\documentclass[aps,twocolumn]{revtex4-1}
\usepackage{amsmath,amssymb}
\usepackage{slashed}
\usepackage{graphicx}
\usepackage{bm}
\usepackage[T1]{fontenc}
\usepackage[utf8]{inputenc}
\usepackage{gauss} 
\usepackage[top=1.0in,bottom=1.0in,left=1.0in,right=1.0in]{geometry}
\usepackage[colorlinks,linkcolor=blue,citecolor=blue]{hyperref}
\usepackage{comment}
\usepackage{ytableau}
\newcommand{\be}{\begin{equation}}
\newcommand{\ee}{\end{equation}}
\newcommand{\bea}{\begin{eqnarray}}
\newcommand{\eea}{\end{eqnarray}}
\newcommand{\ba}{\begin{eqnarray}}
\newcommand{\ea}{\end{eqnarray}}

\begin{document}

\title{Pentaquarks  on the light front and their mixing to baryons
}

\author{Nicholas Miesch}
\email{nicholas.miesch@stonybrook.edu}
\affiliation{Center for Nuclear Theory, Department of Physics and Astronomy, Stony Brook University, Stony Brook, New York 11794--3800, USA}

\author{Edward Shuryak}
\email{edward.shuryak@stonybrook.edu}
\affiliation{Center for Nuclear Theory, Department of Physics and Astronomy, Stony Brook University, Stony Brook, New York 11794--3800, USA}

\author{Ismail Zahed}
\email{ismail.zahed@stonybrook.edu}
\affiliation{Center for Nuclear Theory, Department of Physics and Astronomy, Stony Brook University, Stony Brook, New York 11794--3800, USA}

\begin{abstract} 
In previous papers we developed the light front formulation for Hamiltonians and wave functions (WFs) for mesons and baryons, with both confinement and chiral symmetry breaking. For baryons limited to the lowest Fock component with three quarks, the longitudinal WF is valued in
an equilateral triangle with momentum fractions $x_i,i=1,2,3$. The WF was developed both numerically and using a basis function that diagonalizes the Laplacian with Dirichlet boundary conditions. 
In this paper we extend this analysis to $n$ quark states, and specialize to pentaquarks ($n=5$). We
determine their masses and WFs,
and  address the  mixing between baryons and pentaquarks, the issue  central to understanding the observed antiquark sea of baryons.
 \end{abstract}
\maketitle

\section{Introduction}
This paper is part of a series devoted to bridging the gap between hadronic spectroscopy and partonic observables. In the original paper~\cite{Shuryak:2021fsu} we showed how the quark model for hadrons with constituent quarks and confinement, can be extended to the light front. One chief difference with other formulations (e.g.\cite{Vary:2009gt,Jia:2018ary}) is in the simplicity of the Hamiltonian as suggested by QCD scalar confinement, which is
diagonalized using the CM free (Jacobi) variables. This procedures  is routinely used  in hadronic spectroscopy in the CM frame.

We analyzed the  mesons in~\cite{Shuryak:2022wtk}, and then the baryons in~\cite{Shuryak:2022thi,Shuryak:2022wtk}. In both descriptions, the Hamiltonian included both the kinetic energy and confinement, quantized in the momentum representation, with the hyperfine interactions treated in perturbation. The  light cone WFs depends on the (Bjorken-Feynman) momentum fractions  inside the 3d cube $x_1,x_2,x_3\in [0,1]^3$, cut by the momentum normalization condition $x_1+x_2+x_3=1$. As shown in  Fig.\ref{fig_cube}, this leads to quantum mechanics on a 2d equilateral triangle
(regular 2-simplex called $A_2$). For completeness, some of the details of this procedure  will be briefly recalled  in the section~\ref{sec_bar} below. In this paper we generalize this approach to pentaquarks. The forward part of the wave functions is defined at regular 4-simplex called $A_4$. We study WFs on it, both by approximate method (Ritz variational)
and exact analytic one based on sum of many interfering waves.

Several of our recent papers developed a novel approach 
to finding the WFs antisymmetric  under quark permutations, as required by Fermi statistics.
 This method was used to analyze baryons in P-shells~\cite{Miesch:2023hvl} and D-shells \cite{Miesch:2025vas}, tetraquarks in \cite{Miesch:2023hjt} and even multi-quark states with 6, 9 and 12 quarks.
In all those cases the method is based on {\it a brute force} representation
of the permutation groups $S_N$ generators and their diagonalization,
first in special ``good basis" notations, and then translating to standard monom representation. Since in 
the pentaquark case discussed
the ``monom space" is fairly large, ${\cal O}(10^6)$, we relied heavily on Wolfram Mathematica. Needless to say, that the method was checked to reproduce the standard  group representations for few quark systems. The latter becomes increasingly untractable for higher quark systems, as the mixed representations mushroom, hence the need of a more 
advanced approach.

 In a recent study~\cite{Miesch:2025wro}, we analyzed (light quark) $pentaquarks$ in this framework. More specifically, we derived the explicit WFs for their S- and P-shell states, 
which are needed for the  baryon-pentaquark mixing. This is key to solving long-standing puzzles, such as  the flavor asymmetry of the antiquark sea, and the strong orbital motion in the nucleons. The present paper follows~\cite{Miesch:2025wro} closely,
by extending the description of pentaquarks to the light front (LF).

The paper is organized as follows. We start with \ref{sec_LF} where 
the main ideas and technical steps encountered are recalled, by setting 
quark models on the light front. These are then applied to light pentaquarks
$qqqq\bar q$ in section \ref{sec-penta}. The representations of the permutation group $S_4$, and the explicit construction of the  orbital-color-spin-flavor wave functions for S-shell and P-shell pentaquarks, are carried in section \ref{sec_permut}. The P-wave (negative parity) states are 
important for addressing the $\bar q q$ admixture in  $qqq$ baryons,
as discussed in section \ref{sec_mixing}. The relation between the
results obtained and the properties of the antiquark sea of the nucleon,
are discussed in section \ref{sec_sea}. The paper is summarized 
in section \ref{sec_summary}. 

Most of technical content of the paper is delegated to Appendices.
Appendix A recalls the quantization method for baryons and other hadrons on the light front (LF), in particularly their relation to the well known problem
of $n$ fermions on a circle. Appendix B considers a variational 
calculation of the longitudinal pentaquark  wave function. Appendix C derives
an exact eigensystem for the Laplacian on the $A_4$ simplex, as needed for pentaquarks. Appendix D considers the  large $N$ limit, in which pertinent 
simplexes can be approximated by spheres. 

\section{Quark models  on the light front} \label{sec_LF}

Quark models, including confining potentials and spin-dependent forces,  have for decades provided a reasonably good explanation of hadronic
spectroscopy. They are naturally formulated
in the center of mass (CM) frame. One may wander why
people, us included, have put efforts to re-derive
them using LF kinematics. 

One reason is to get a more direct connection to the relativistic form-factors and 
``partonic observables" (PDF, TMD, GPD etc). Another, is that in the LF
formulation one can treat light and heavy quark systems
on equal footing, the nonrelativistic approximation is not needed. Finally,
what looks like static $\pi,\sigma$ ``clouds" around hadrons in
their rest frame, under a boost they give rise to the $\bar q q$ ``sea", 
or multi-quark components of the wave functions.
The complex  nature of the boost operator is what forced the development
of hadronic models  directly on the LF.  These models usually pausit
some Hamiltonian, and then solve it to account for the
full spectroscopy, with masses and wave functions.

\subsection{Confined N-quarks in 1+1 dimensions}
For simplicity, consider first the dimensionally reduced
longitudinal LF Hamiltonian for $N$ confined quarks of equal mass,
\bea
\label{HLF4}
H_{LF,L}=\sum_{i=1}^N\bigg(\frac{m_Q^2}{x_i}
+2\sigma_T  \bigg|\frac{i\partial}{\partial x_i}\bigg|\bigg)
\eea
The corresponding WFs for N-quarks are solutions to 
\bea
\label{HLF5}
\sum_{i=1}^N\bigg(\frac{m_Q^2}{x_i}
+2\sigma_T  \bigg|\frac{i\partial}{\partial x_i}\bigg|\bigg)\varphi_n[x]
=
M_n^2 \varphi_n[x]
\eea
giving mass squared of specific states.

Note that this Hamiltonian already includes two main
non-perturbative phenomena in QCD.
The first term is the kinetic energy, it includes
effective ``constituent" quark masses.
The second, proportional to the string tension $\sigma_T$, describes quark confinement. It will be solved in the momentum representation, so the
coordinates are  derivative of  momenta.

\begin{figure}[t!]
    \centering
    \includegraphics[width=6cm]{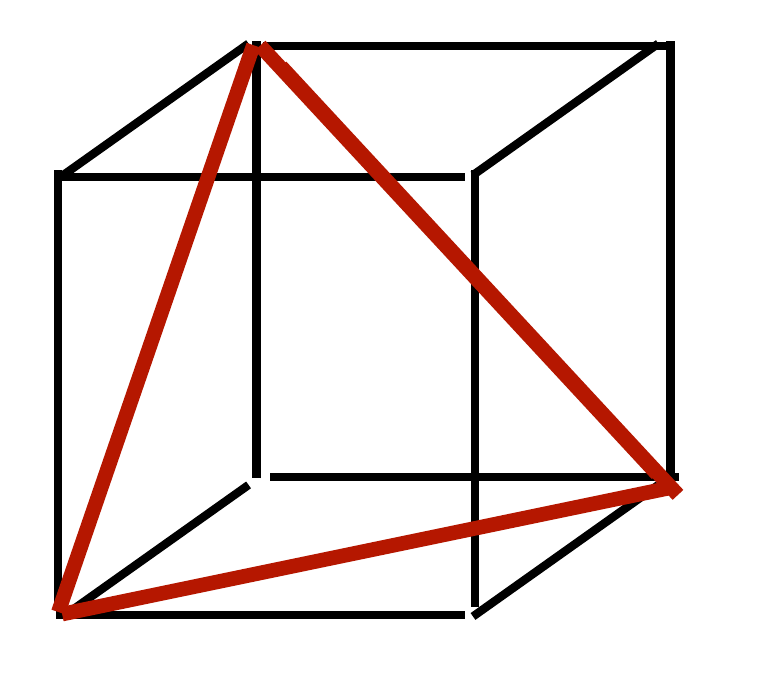}
    \caption{equilateral triangle formed by cube $x_1,x_2,x_3\in [0,1]^3$ cut by momentum normalization condition $x_1+x_2+x_3=1$}
    \label{fig_cube}
\end{figure}

 
 The variables $x_i$ are Bjorken-Feynman momentum fractions
 for  N identical particles moving  in
 a box  $0\leq x_i\leq 1$ and subject to the momentum constraint \bea
\label{SUMX}
X=\sum_{i=1}^Nx_i=1
\eea
 The  condition (\ref{SUMX}) makes the physical support
 of the problem non-trivial. It is an $\tilde N$-simplex with
one dimension lower $\tilde  N=N-1$. 
For baryons $N=3$, the domain is an equilateral triangle,
see Fig.\ref{fig_cube}, and our solution for quantum mechanics on it was developed earlier. 
For larger $N$
 the domain is the N-dimensional generalization of an equilateral triangle. For  $N=4$ it is a tetrahedron or 3-simplex, and for pentaquarks $N=5$ it is the 4-dimensional
 5-simplex $A_4$. A pentachoron with 5 corners, 10 edges and 10 triangular faces  also referred to as a  ``starfish".
 
\begin{figure}[h]
\begin{center}
\includegraphics[width=6cm]{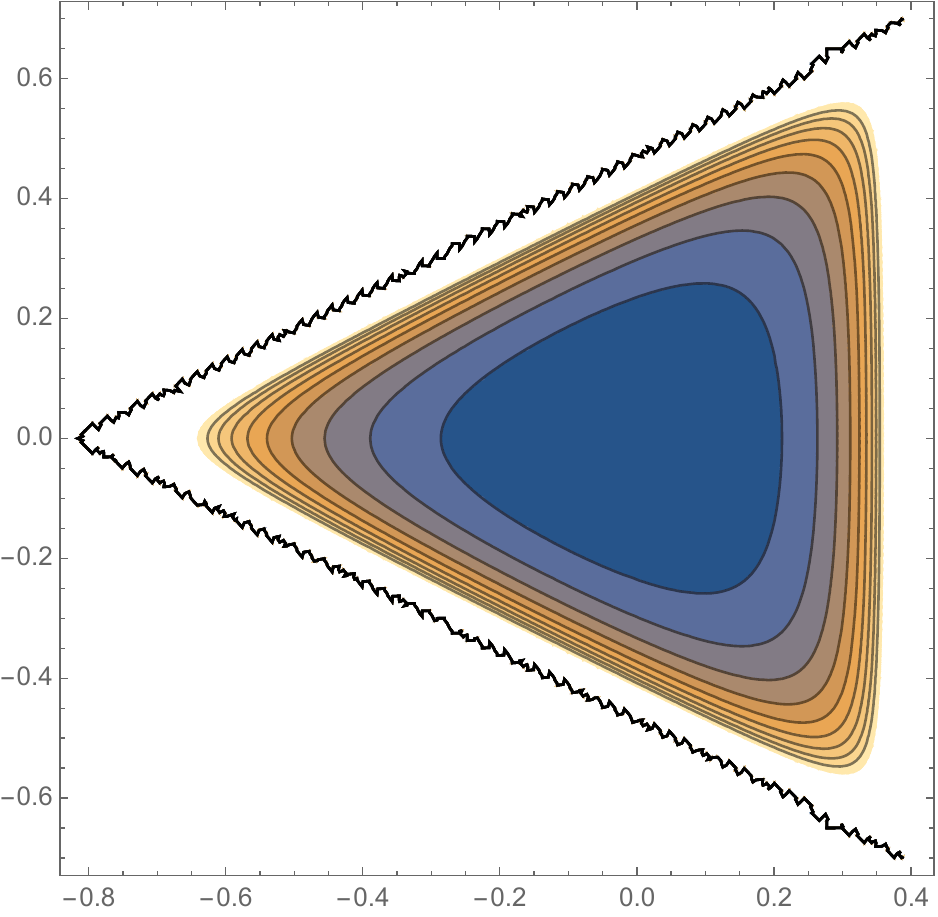}
\includegraphics[width=1cm]{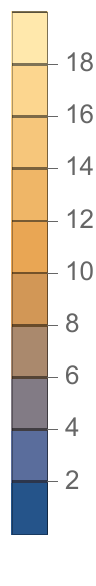}
\caption{Effective potential for equilateral triangular domain for $N=3$.}
\label{fig_V_in_Jacobi}
\end{center}
\end{figure}

  The singular kinetic energy term forces the wavefunction to vanish at these boundaries (Dirichlet boundary condition). The shape of the effective ``cup potential" is shown in Fig.\ref{fig_V_in_Jacobi} for the baryons.

To solve this problem, we proceed in three steps. First, we unwind the 
square roots  by using the einbein trick 
\bea
\label{CONF2}
\sum_{i=1}^N\bigg|\frac{i\partial}{\partial x_i}\bigg|&=&\sum_{i=1}^N\frac 12\bigg(\frac 1{e_{iL}}+e_{iL}\bigg(\frac{i\partial}{\partial x_i}\bigg)^2\bigg)\nonumber\\
&\rightarrow&\frac 12\bigg(\frac 3{e_{L}}+e_{L}\sum_{i=1}^N\bigg(\frac{i\partial}{\partial x_i}\bigg)^2\bigg)
\eea
and  assume equal $e_{iL}=e_L$ at the extrema, in the steepest descent approximation.
Second, we diagonalize the free Laplacian after isolating the center of mass coordinate, using normal mode coordinates  (analogue of Jacobi coordinates) see below. Third, we
minimize the energies with respect to $e_L$.

The Hamiltonian and wave functions use the momentum representation and Jacobi coordinates (so, unlike
the work carried by Vary's group, we do not have any issue with
the center-of-CM motion, which is set to zero from the onset.). 

There are two methods to do quantum mechanics for such
a Hamiltonian: \\ (i) One, derived  earlier for a triangle,
is to work out the complete set of eigenfunctions for 
the Laplacian (the second term), and then represents the second
term as a matrix. In some truncated form it can be
diagonalized.\\
(ii) For a triangle, one can numerically solve the Schrodinger equation, with the laplacian and a cup potential included. For the reader convenience, the main points in this approach 
can be found in Appendix~\ref{sec_bar}. Recently, for pentaquarks we were able to numerically diagonalize
this Hamiltonian by using a set of determinantal wavefunctions~\cite{PENTAX}. Here we will develop a variational approximation, as well as a simplex-motivated derivation.

\section{Pentaquarks on the light front } \label{sec-penta}
\label{sec_penta}

The first step in conventional spectroscopy (in the CM frame)
is the selection of appropriate coordinates, excluding
the fictitious motion of the center of mass. In the momentum
representation and for momentum fractions, this is captured by 
the longitudinal momentum constraint (\ref{SUMX}). In both cases
the problem was solved for us long ago, by
choosing Jacobi coordinates. For baryons (N=3) they are
traditionally called $\rho,\lambda$
 \ba \label{eqn_Jacobi_x}
x_1&=&(\sqrt{6}\lambda+3\sqrt{2}\rho+2X)/6 \nonumber \\
x_2&=&(\sqrt{6}\lambda-3\sqrt{2}\rho+2X)/6 \nonumber \\
x_3&=&(-\sqrt{6}\lambda+X)/3 \ea
Note that Laplacian in Jacobi coordinates is 
\bea
\label{SUMXX}
\sum_{i=1}^3\bigg(\frac{i\partial}{\partial x_i}\bigg)^2=
\bigg(\frac{i\partial}{\partial\lambda}\bigg)^2+\bigg(\frac{i\partial}{\partial\rho}\bigg)^2+3 \bigg(\frac{i\partial}{\partial X}\bigg)^2
\eea
and because the total momentum does not change, only the two first
terms remain. So, without a cup potential, the problem 
thereby is reduced to  two free particles confined
to the equilateral triangle box. Our initial study addressed this problem.

We follow up on this work by considering five partons,  subject to the longitudinal constraint
  $\sum_{i=1}^5 x_i=1$. Using the (modified) Jacobi
coordinates $\alpha,\beta,\gamma,\delta$ amounts to
\begin{eqnarray}
 \label{eqn_Jacobi_5}
x_1 &=& (6 + 15 \sqrt{2} \alpha+ 5 \sqrt{6} \beta+ 
    5 \sqrt{3} \gamma+ 3 \sqrt{5} \delta)/30  \nonumber \\
x_2 &=&  (6 - 15 \sqrt{2} \alpha+ 5 \sqrt{6} \beta+ 
    5 \sqrt{3} \gamma+ 3 \sqrt{5} \delta)/30, \nonumber \\
x_3 &=&  (6 - 10 \sqrt{6} \beta+ 5 \sqrt{3} \gamma+ 
    3 \sqrt{5} \delta)/30, \nonumber \\
x_4 &=& 1/10 (2 - 5 \sqrt{3} \gamma+ \sqrt{5} \delta), \nonumber \\
x_5 &=& 1/5 - (2 \delta)/\sqrt{5} \\
\end{eqnarray}
The corresponding 4-dimensional domain 
 (similar to Fig.\ref{fig_cube} but for obvious reason not plotted) is 
  a simplex with 5 pointed corners called $A_4$.
   Five corners, as always, correspond to
 one of the momentum fraction  carried by a single parton $x_i=1$, with all others null. Their explicit locations are
 \ba a_1 &=& \{1/\sqrt{2}, 1/\sqrt{6}, 1/(2 \sqrt{3}), 1/(2 \sqrt{5})\} \nonumber\\
a_2 &=& \{-(1/\sqrt{2}), 1/\sqrt{6}, 1/(2 \sqrt{3}), 1/(2 \sqrt{5})\}
\nonumber\\
a_3 &=& \{0, -\sqrt{2/3}, 1/(2 \sqrt{3}), 1/(2 \sqrt{5}) \} \nonumber\\
a_4 &=& \{0, 0, -(\sqrt{3}/2), 1/(2 \sqrt{5}) \} \nonumber\\
a_5 &=& \{0, 0, 0, -(2/\sqrt{5}) \}
 \ea
All have norms  $4/5$ and all 10 mutual products are equal to $-1/5$ that angles between them $\theta_{i>j}$ are all the same, $cos(\theta_{i>j})=-1/4$.

There are 10 segments connected two corners $a_i,a_j,i\neq j$,
called ``edges". For example, the previous formulae show that $a_4$ to $a_5$ belong to $\alpha=\beta=0$ plane. Across them there are 
``faces" with three corners $a_i,a_j,a_k, \, i\neq j \neq k$,
(e.g. 1,2,3). Those are 2-dimensional equilateral triangles.
``Across" means that centers of edges ($(a_i+a_j)/2$ and centers
of complimentary faces $(a_i+a_j+a_k)/3$ are connected by lines going though
the global center $\alpha=\beta=\gamma=\delta=0$.

Our task now is to do quantum mechanics on this simplex. We recall that by a Hamiltonian on the LF,
we mean $M^2=2P^+P^-$.
Using the same einbein trick, one can get rid of the square roots in the confining terms, so
that the linear confining part of the Hamiltonian 
is traded for a quadratic form in coordinates. In  momentum representation it is a Laplacian. The transverse coordinates can also be replaced by derivatives $r^2_{i\perp}=-\partial/\partial_{k_{i\perp}^2}$, and so our LF Hamiltonian is 
\bea\label{eq:LFHami}
H_{LF}&=&\sum_{i=1}^5\frac{k^2_{i\perp}+m_Q^2}{x_i}
\nonumber\\
&+&\sigma_T\left(5a+\frac{1}{a}\sum_{i=1}^5((i\partial/\partial x_i)^2-M^2 (\partial/\partial_{\vec{k}_{i\perp}})^2)\right) \nonumber\\
\eea
Note that we neither have attractive Coulomb forces nor a (negative)
constant, which are normally included in the Cornell-type potentials.
Therefor we would not compare the absolute eigenvalues of this
Hamiltonian, but rather its differences and the wave functions.

The first term (former kinetic energy in LF) is rewritten by adding and subtracting  mean terms twice
\bea
&&\sum_{i=1}^5\frac{k^2_{i\perp}-\langle k^2_{i\perp} \rangle}{x_i} 
 \nonumber \\
&&+ \langle k^2_{i\perp} +m_{Qi}^2\rangle\bigg(
\sum_{i=1}^5\frac{ 1}{x_i} - 25\bigg) \nonumber \\ 
&&+25(\langle k^2_{\perp} \rangle+m_Q^2)
\eea
The first term is referred to as the ``non-factorization potential", the second is ``the cup potential"
and the last is just a constant.
The cup potential is a function of only momentum fractions $x_i$ on the simplex. It 
is near zero in the middle of the simplex, $\alpha\approx \beta\approx \gamma\approx\delta\approx 0$  and increases near all boundaries for which $x_i(\alpha,\beta,
\gamma,\delta)\rightarrow 0$. To avoid divergences at the simplex boundaries,
the wave functions need to vanish there. More specifically,
we will discuss only the cases with finite $\psi V \psi $, which requires vanishing $\psi(\rm near \,boundary)\sim (A-A_{boundary})^p$ with power $p>1/2$. 

If the ``non-factorization potential" is small and neglected, then this Hamiltonian is a sum
of transverse and longitudinal parts, and the WFs 
factorize  into parts depending on $k_{\perp}$ and the longitudinal momentum fractions $x_i$. The transverse part is approximated by an oscillator
$$ H_\perp=5\sum k_{i\perp}^2- {\sigma_T M^2\over a}\sum \big(\partial/\partial k_{i\perp})^2)\big)  $$ 
so e.g. its ground state wave function in 1d  is 
\bea
\label{eqn_osc_wf}
&&\psi_0\sim exp(-k^2_\perp\beta_{osc}^2/2), \nonumber\\
&&\beta_{osc}=\big({5a\over 25\sigma_TM_q^2}\big)^{1/4} 
\eea
with  energy $E_0=5^{3/2} M_q \sqrt{\sigma_t/a}\approx 0.67\, GeV^2$ for $M_q=0.35\, GeV, \sigma_T=0.42^2 GeV^2, a\approx 6$ (as follows after minimization in $a$).
The once-excited level has  $\psi_1\sim k_\perp exp(-k^2_\perp\beta^2/2)$ and energy $E_1=3E_0$. Generalization
to 2d oscillator is obvious. 

In a separate paper \cite{PENTAX} we discussed a well known and somewhat related problem, that of $N$ fermions on a circle. Since fermions cannot pass each other in 1d, their coordinates
$s_i$ are ordered $s_1<s_2<...s_N<s_1+1$. Furthermore, one may
fix the CM by a condition $\sum_i s_i=0$, so its physical domain
becomes a 4-dimensional simplex. However, it is important to note  that it is 
similar (but $different$) from the symmetric  simplex\footnote{In Wolfram Mathematica there is a command $starfish=Simplex[a_1,...a_5]$ creating simplex between given number of points. It can be used in integrals 
as the domain.} $A_4$ for the pentaquark problem we are interested in.
The fermions-on-circle problem is useful because it 
has known analytic solutions. Its WFs have the form of Slater determinants~\cite{Girardeau:1960cnk}  containing four momenta $n_{21},n_{31},n_{41},n_{51}$ which should all be nonzero and different,
so that Slater determinant does not vanish. Naturally, for the ground state one should select the ``occupied Fermi sphere" set, taking their values to be the lowest possible, namely $-2,-1,1,2$ (recall that momentum zero is already included).
The Laplacian eigenvalue is then proportional to the sum of their squares $0^2+2*1^2+2*2^2=10$. Opening the determinant one finds 
sum of many waves, a very lengthy expression. Yet it can be 
 simplified  to the product of 10 sinusoidal terms
 for each $i\neq j$ pairs
\be \label{eqn_prod_sin}
\psi_{sin} \sim \prod_{i\neq j} sin\bigg( {s_i-s_j \over 2}\bigg) \ee
While the 4-d simplex of fermions-on-the-circle problem is $not$ the same
as we need for pentaquarks, it can be mapped to it. Unfortunately, this map changes 
 the Laplacian  into a different 
elliptic operator. Therefore, nice classic fermions-on-circle problem 
does not solve the one we need. Yet its set of states can be used as an orthonormal basis, see our companion paper~\cite{PENTAX}.

Since we will discuss  the lowest S-shell and the next
(negative parity) P-shell states only, we will use the variational approximation.
For the ground state
$\phi_0(\alpha,\beta,\gamma,\delta)$ we studied
 two different trial functions, one motivated by the 
 fermion-on-circle problem (\ref{eqn_prod_sin})
 and another one with minimal enforcement
 of the boundary conditions. 
The first is the product of sinusoidal functions along 
  ten axes, connecting the centers of the  ``edges" to the centers of the ``faces".
  Their arguments are such that  they vanish at the above-mentioned ends.
 \be \Psi_{sin}=\prod_{i\neq j}sin\bigg({2\pi \over 5}+2\pi (\rm coor\cdot n_{ij}) \bigg), \ee 
where the coordinate vector $coor=(\alpha, \beta,\gamma,\delta)$
is multiplied by 10  unit vectors $n_{ij}$ pointing to the centers
of all 10 edges. 
 Needless to say, it is symmetric
  for all quark permutations.

\begin{figure}[h!]
    \centering
    \includegraphics[width=7cm]{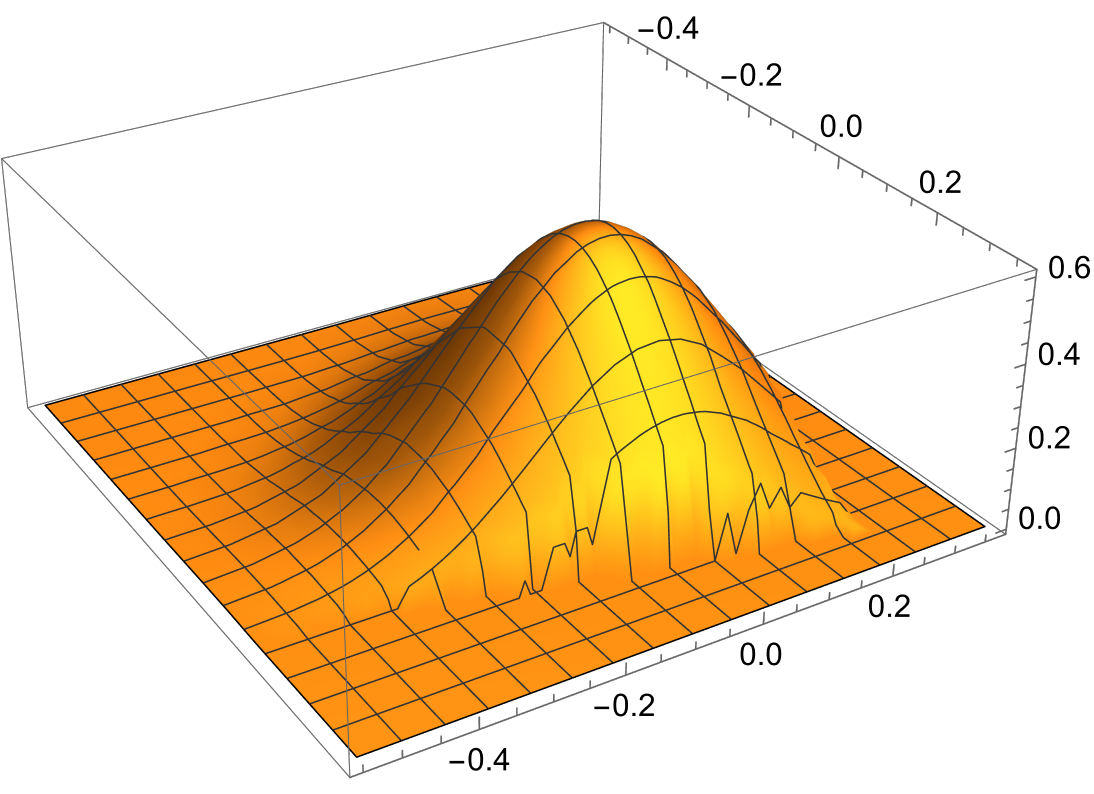}
    \caption{The shape of the  $\Psi 0^2$, at $\alpha=\beta=0$   at the  $\gamma,\delta$ plane. Note that
    this plane contains two corners $a_4,a_5$, 
    but not the opposite center of the face 123: thus
    it does not sit on an equilateral triangle.}
    \label{fig_penta_vary_WF1}
\end{figure}
 
Another (and even simpler) trial function
 is the $product$ of linear functions $x_i,i=1..5$ (from (\ref{eqn_Jacobi_5}))
\be \label{eqn_ansatz_faces}
\Psi_{faces}=\prod_{i=1..5} x_i(\alpha,\beta,\gamma,\delta)
\ee
It also  
 enforces the Dirichlet boundary conditions. We found that 
 not only it approximates the lowest eigenstate of
 the Laplacian, but with the inclusion of the ``cup potential"
 it even becomes closer to it. This second trial function,
 appropriately modified, also gives an approximation to
 the negative parity P-wave WF.
 The details are in Appendix \ref{sec_var}.


Another technical point, discussed in
 Appendix \ref{sec_large_N}, is a study of the problem using $spheres$, to which the N-simplexes tend at large $N$. While the average values of the Laplacians and
Laplacian+$V_{\rm cup}$
 may
differ substantially, the corresponding wave functions are 
actually much closer. Our variational ground state wave function
is a reasonable approximation to 
the (forward part) of the pentaquark wave function. This will be confirmed by 
a direct comparison to the exact diagonalization carried in~\cite{PENTAX}.

\begin{figure}
    \centering
    \includegraphics[width=0.7\linewidth]{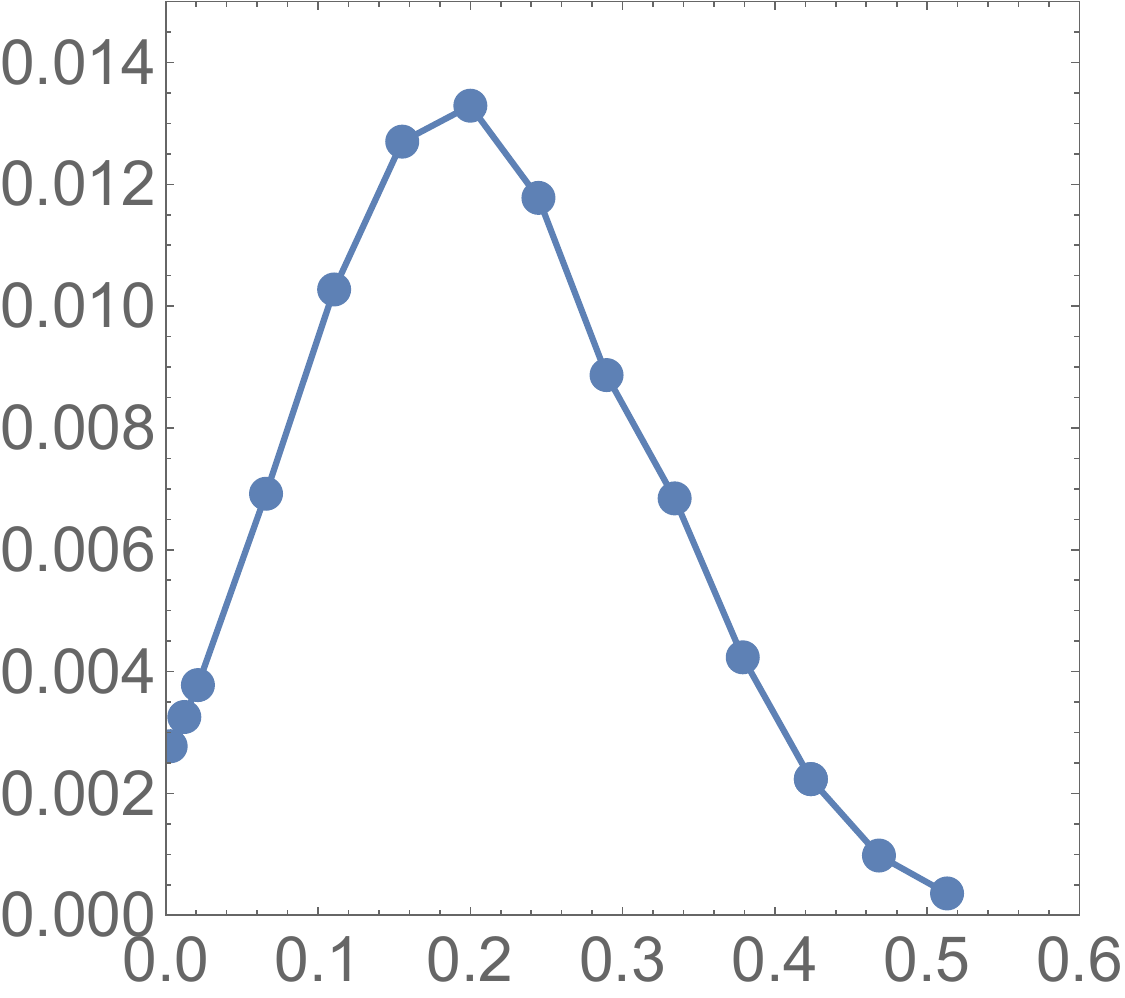}
    \includegraphics[width=0.7\linewidth]{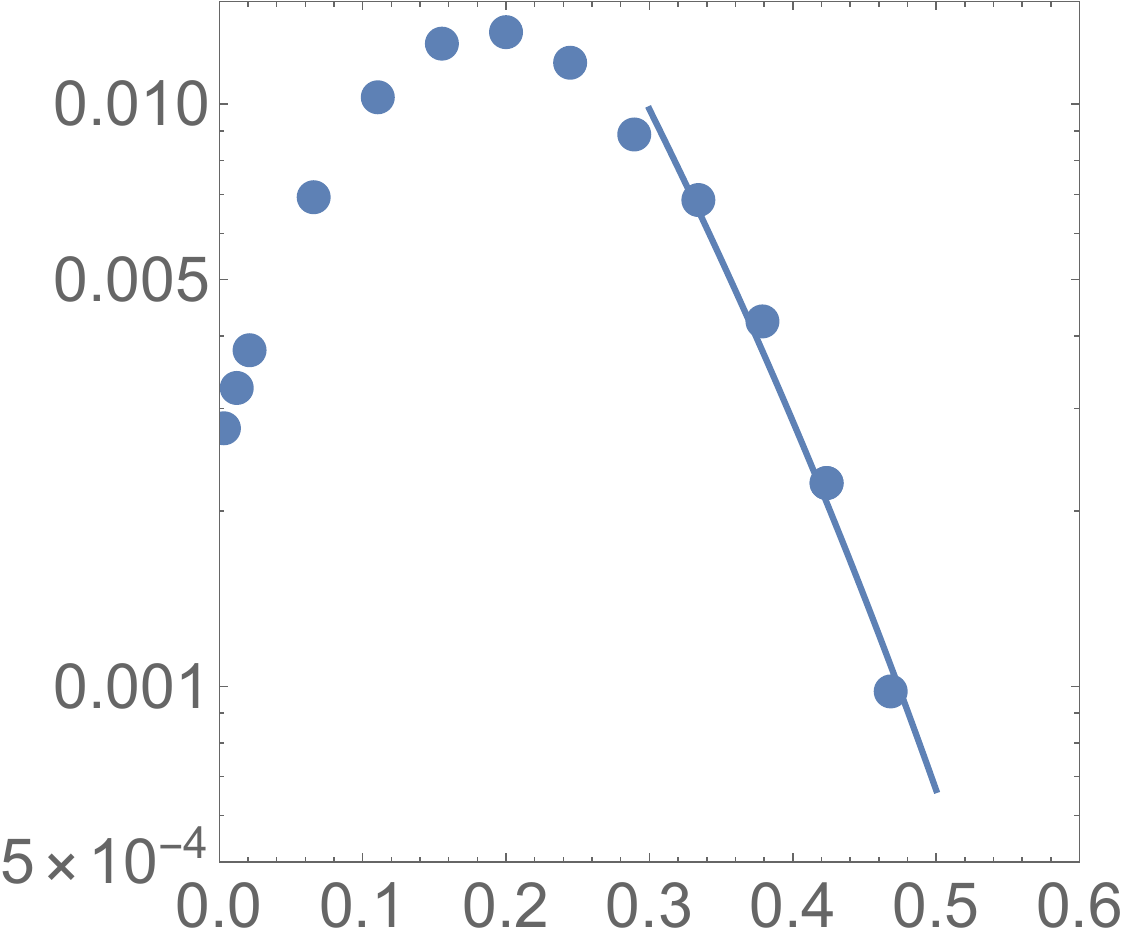}
    \caption{The antiquark PDF in pentaquarks versus the momentum fraction $x$ (arbitrary units). The upper and lower plots show the same distributions, in linear and logarithmic plots.
    The lower plot is used to magnify the larger $x$ region, with a fit to a form $\sim (1-x)^8$.}
    \label{fig_penta_pdf}
\end{figure}

The pentaquark wave function can provide a full description of the quantum states,
with their associated observables. For example, the squared longitudinal part of the WF (depending on momentum fractions) is the PDF, the probability of one particle to have definite  momentum fraction, with all others integrated out.  Also, the antiquark PDF $P(x_{\bar q})$  is obtained by integration of the wave function squared over $\alpha,\beta,\gamma$ variables, as a function of  the last Jacobi coordinate $\delta$. 
$$ PDF_{\bar q}(\delta)=\int_{\alpha,\beta,\gamma} |\Psi|^2 $$
The results are shown in Fig.\ref{fig_penta_pdf}. 
As expected for pentaquarks, the maximum is close to
$x_{mean}=1/N=1/5$. A nontrivial prediction is that it gets very small
for $x>0.5$, and its hard edge behavior is $\sim (1-x)^8$.

\section{Orbital-color-spin-flavor wave functions on LF from the permutation group $S_4$} \label{sec_permut}
In the CM frame the ground state is approximately spherical,
justifying the ``hyperdistance" approximation.
In the LF  spherical symmetry is absent, and the  WF is approximated by
a product of two parts, depending on the transverse and longitudinal momenta, respectively. 

So far we discussed 
the latter for the ground state,
as a function of the hyperdistance, which is symmetric under
quark permutations. There is also
a nontrivial part in the WFs that carry the 
color-spin-flavor indices. The only difference
is that in the traditional (CM frame) spectroscopy   there is 
$spherical$ symmetry $O(3)$, while on the LF there is only $axial$ symmetry $O(2)$. Therefore, the usual classification based on quantum numbers $J,L,S$ is now unavailable. Instead, we should use only 
their longitudinal components $J_z,L_z,S_z$.  

Considering states with nonzero orbital
momentum $L\neq 0$, as we will do below, 
one finds further differences between
the spectroscopy in the CM and the LF. 
In the former case   the  wave function with fixed total angular momentum and projection $J,J_z$ (e.g.$J=1/2,J_z=1/2$) are built by a standard sum weighted by Clebsch-Gordon coefficients
$$ \sum_{Lz} \big(^{1/2,1/2}_{L,Lz,S,1/2-Lz} \big)
Y_{L,Lz}(\theta,\phi)|S,Sz \rangle $$
corresponding to $\vec L+\vec S=\vec J$.
In contrast, on the LF  the $L,S$ quantum numbers are absent. In particular, 
for the negative parity P-shell there are three values of $L_z=-1,0,1$, to be combined with
six possible values of $S_z=-5/2,-3/2,...,5/2$.
The only global
quantum number is  their sum $J_z=L_z+S_z$.
Mixing of all states with the same $J_z$ happens,
e.g. defined by the spin-dependent terms in the Hamiltonian.
 Components with $L_z=1$ and $L_z=-1$ add one quantum of transverse
oscillator, and the last  $L_z=0$ contains
an extra linear coordinate
$z=-i{\partial \over \partial p_z}$.

\subsection{$L=0$ pentaquarks on LF}
In our previous paper \cite{Miesch:2025wro} we derived fully antisymmetric
WFs for pentaquarks, belonging to  S- and P-shells. The procedure was based on representations
of the permutation group $S_4$.
Now we need to  apply similar procedures on the LF. 

In the S-shell (L=0)
the only changes are in the spin classification. For 5 quarks
there are six choices, $S_z=-5/2,3/2...5/2$. In the extreme 
$S_z=5/2$ case, the spin WF  is a single symmetric monom $\uparrow \uparrow\uparrow\uparrow\uparrow$. Therefore Fermi statistics must be satisfied by (still nontrivial) $color\otimes flavor$
wave functions. Those were already found in \cite{Miesch:2025wro}
for $I=1/2$ and $I=3/2$ (there is no state with $I=5/2$.)

With the new ``axially symmetric" $I,S_z$ classification the number of states and
permutation generators are given by different matrices, yet the procedure is generally the same.
Let us jump to a table  giving the number of appropiate pentaquark antisymmetric states, with fixed values of isospin and spin projection $I,S_z$:

\begin{table}[h!]
    \centering
    \begin{tabular}{|c|c|c|c|} \hline
  $I/S_z$    & 1/2 & 3/2   & 5/2  \\  \hline
 -5/2  & 1 (15) & 1 (12) & 0\\
 -3/2 & 4 (75) & 4 (60) & 1 (15)\\
 -1/2 & 7 (150) & 7 (120)  & 2 (30)\\
  1/2 &  7 (150) & 7 (120)  & 2 (30) \\
  3/2 &4 (75) & 4 (60) & 1 (15)    \\
  5/2 & 1 (15)  & 1 (12)  & 0 \\ \hline
    \end{tabular}
    \caption{Table of the pin projection $S_z$ versus isospin $I$ 
    of antisymmetric pentaquark states
    with $L=0$. The numbers in brackets refer to the dimensions of the ``good basis",
    in which permutation matrices are calculated and diagonalized.
   .}
    \label{tab_L0}
\end{table}


It is instructive to explain how this table is  related to the one we had
derived in the CM frame in \cite{Miesch:2025wro}. 
Let us s count the total number of states with $I=1/2$. The Table in
\cite{Miesch:2025wro} would give a total number of states
 $$2*3+4*3+6*1=24$$ 
where the first numbers are the number of components, $2S+1$ for fixed
spin $S$. Now there is no spherical symmetry and the total spin is not used. 
In the current Table the number of states with $I=1/2$ (the first column) looks simply as: $$ 2(7+4+1)=24 $$
where the 2 in front stands for doubling due to negative $S_z$. 
These are the same states but counted differently. Different
$2S+1$ orientations of spin-$S$ states related by rotations
are no longer related, as the only rotations allowed on the LF
are the axial ones, preserving $S_z$. 

Let us introduce the (relativistic correction) color-spin Hamiltonian
which is based on products of color and spin matrices for each particle pair
\be \label{ean_H_lambdalambdasigmasigma}
H_{\lambda\lambda \sigma \sigma}=-\sum_{i>j} V_{\lambda\lambda \sigma \sigma}^{ij}(\vec \lambda_i\vec \lambda_j)(\vec \sigma_i\vec \sigma_j)\ee
and calculate the Hamiltonian matrix including all states. Here we are substituting the average value of the potential, taken with
the main symmetric WF, ignoring the back reaction
on the spins. 
All matrix elements are nonzero $O(1)$ numbers and look complicated, but after its diagonalization we found that its eigenvalues reproduce
exactly the same values as we got using the ``spherical" classification
in \cite{Miesch:2025wro}. In particular, the 7 states with $S_z=1/2,I=1/2$ have energies 
\ba E_n &=&4.6667,  3.0000, 1.4415, -2.7749,\nonumber \\
&-&0.33333,- 3.3333, -3.3333   \ea
all of which have been reported there. Four states with 
$S_z=3/2,I=1/2$ have energies 
\be E_n= 3.000000, -0.3333333,-3.333333, -3.333333\ee 
Repetitions of the same values as for $S_z=1/2$ is required by (now ``hidden") $O(3)$ symmetry, not used but still present in the Hamiltonian.

\subsection{$L=1$ pentaquarks on the LF}
Moving up, to what used to be called $P$-shell states in the CM frame,
we recall that negative parity requires  some linear combination of coordinates, either transverse or longitudinal, times some functions of their squares. Therefore looking for the antisymmetric representation of
the permutation group $S_4$, one 
 needs to work in higher dimensions of monoms, $N_{monoms}=4*3^6*2^5*2^5$. 
For the WFs  as in all other cases, are obtained by explicit
use of the SparseArray command  in Mathematica. 
Unfortunately, those are still way too large
to be presented in print.

The multiplicities of the P-shell pentaquark states were detailed in our previous paper
\cite{Miesch:2025wro}, see Table IV.
On the LF, linear functions of coordinates get changed into linear functions of momenta.  However, the permutation properties defined via Jacobi remains unchanged. Thus the orbital-color-spin-flavor structure
of the states remains the same.  The analogue to the Table IV of \cite{Miesch:2025wro}  is the new Table \ref{tab_L1}.

There are now two types of $L=1$ states: (i) with $L_z=\pm 1$ corresponding to rotations in
the transverse plane,  and (ii) $L_z=0$ case, in which the linear terms are made of the longitudinal momenta.  Because the transverse part of the Hamiltonian 
is the 2-d oscillator, type-1 states contain combinations of transverse momenta $\alpha_1\pm i \alpha_2$ $times$ the ground state Gaussian WFs in transverse momenta.
The same applies to  other Jacobi combinations of $\beta,\gamma,\delta$, with  4 possibilities altogether.  Clearly, there are certain 4-by-4 matrices of permutation generators. 
Those are the same as for longitudinal momenta (fractions) which we call
simply $\alpha,\beta,\gamma,\delta$, omitting the longitudinal index.
Evaluating the Kronecker products of permutation matrices for coordinates, color, spin and flavor, we obtain the global permutations. The  antisymmetric eigenvectors are found by diagonalization.

\begin{table}[h!]
    \centering
    \begin{tabular}{|c|c|c|c|} \hline
  $I/S_z$    & 1/2 & 3/2   & 5/2  \\  \hline
 -5/2  & 3 (60) & 3 (48) & 1 (12) \\
 -3/2 & 14 (300) & 13 (240) & 4 (60)\\
 -1/2 & 27 (600) & 24 (480)  & 7 (120)\\
  1/2 &  27 (600) & 24 (480)  & 7 (120) \\
  3/2 & 14 (300) & 13 (240) & 4 (60)    \\
  5/2 & 3 (60) & 3 (48) & 1 (12) \\ \hline
    \end{tabular}
    \caption{Table of antisymmetric pentastates with spin projection $S_z$ versus isospin $I$,
    with $L=1$. The numbers in brackets are the dimensions of the``good basis", 
    in which the permutation generators are calculated and diagonalized.
   .}
    \label{tab_L1}
\end{table}

Let us mention a  set of energies for $I=1/2,S_z=1/2,L=1  $
states of the color-spin Hamiltonian
\ba E_n &=& 9.05, 9.05, -7.33, 4.67, 4.67, -3.33, \nonumber \\ &&
-3.33, -3.33, -3.33, -3.33, 
-3.33, -3.33,  \nonumber \\ &&
3.00, 3.00, 3.00, -2.77, -2.77, -2.77,
\nonumber \\ &&
1.67, 1.67, 
1.62, 1.62, 1.44, 1.44, \nonumber \\ 
&& 1.44, -0.333, -0.333 \ea
 with repetitions for larger values of $S_z$,
 due to the various components at the same $J$.

A mixing with the proton, $I=1/2$ and spin up $S_z=1/2$ is only possible for 
 pentaquark states with the same quantum numbers.
In the CM frame those include the total angular momentum and its projection,
and mixing with the nucleon is only possible if $J=1/2$. 
The selection of such states is done by standard Clebsch-Gordon
coefficients, combining e.g. $S_z=1/2,L_z=0$ with $S_z=-1/2,L_z=1$
states into unique combination.

However 
on the LF $J$ is not defined and the states with the same $J_z=1/2$,
are treated as independent. The ones with  $L_z=0$
are the type-2 states, and there are 
   27 of them with $S_z=1/2$ as listed in the table. They all have  antisymmetric
  WFs  under  $S_4$. The linear combinations of longitudinal momenta  get combined with appropriate color, spin and isospin structures, as defined in our previous papers.
The type 1 states with $L_z=+1,S_z=-1/2$ 
 are also 27,  with  the transverse momenta occuring linearly 
 in place of the longitudinal ones. 

The type-1 and type-2 states formally do not have the same energy,
as the LF Hamiltonian lacks spherical symmetry.
There is no statement that the full angular momentum operator $\vec J=\vec S+\vec L$ has some fixed values at the resulting WFs. In some cases
it is possible to show that different orientations do lead to the same mass
(e.g. we did it for $\rho$ meson in \cite{Liu:2023fpj}).
 
The ground state transverse WF was given in (\ref{eqn_osc_wf}), and
that of the 1-st excited state is just this one times the coordinate.
The additional energy of these states, compared to the $L=0$, 
comes from one quantum excitation of
some transverse oscillators, with a value
\be M^2(L=1,type\, 1)- M^2(L=0)= 2 \sqrt{5\sigma_T M^2 \over a}
\ee
where $\sigma_T$ is the string tension, and $a$ is the variational parameter of the einbine trick. (Its value at minimum is $a\approx 6$, see \cite{PENTAX}.

\subsection{The shapes of the forward wave functions, for S- and P-shells}

The longitudinal part of the LF Hamiltonian is not an oscillator, 
but a Laplacian (confining) and the irreducible potential $V$.
For 5 quarks those are
defined on the on the pentachoron (``starfish") manifold of $\alpha,\beta,\gamma,\delta$
momentum fractions.

In appendix A we recall  how a similar problem for the 3 bodies (baryons)
and 4 bodies (tetraquarks) can be solved using ``fermions on the
circle. The actual shapes of the wave functions for 5 bodies is determined by a
  variational evaluation in appendix B and by using a more elaborate (but complicated) WFs in Appendix C. Relagating all the details to those Appendices, we would like to present here some results in graphical form.

\begin{figure}[t!]
    \centering 
        \includegraphics[width=0.75\linewidth]{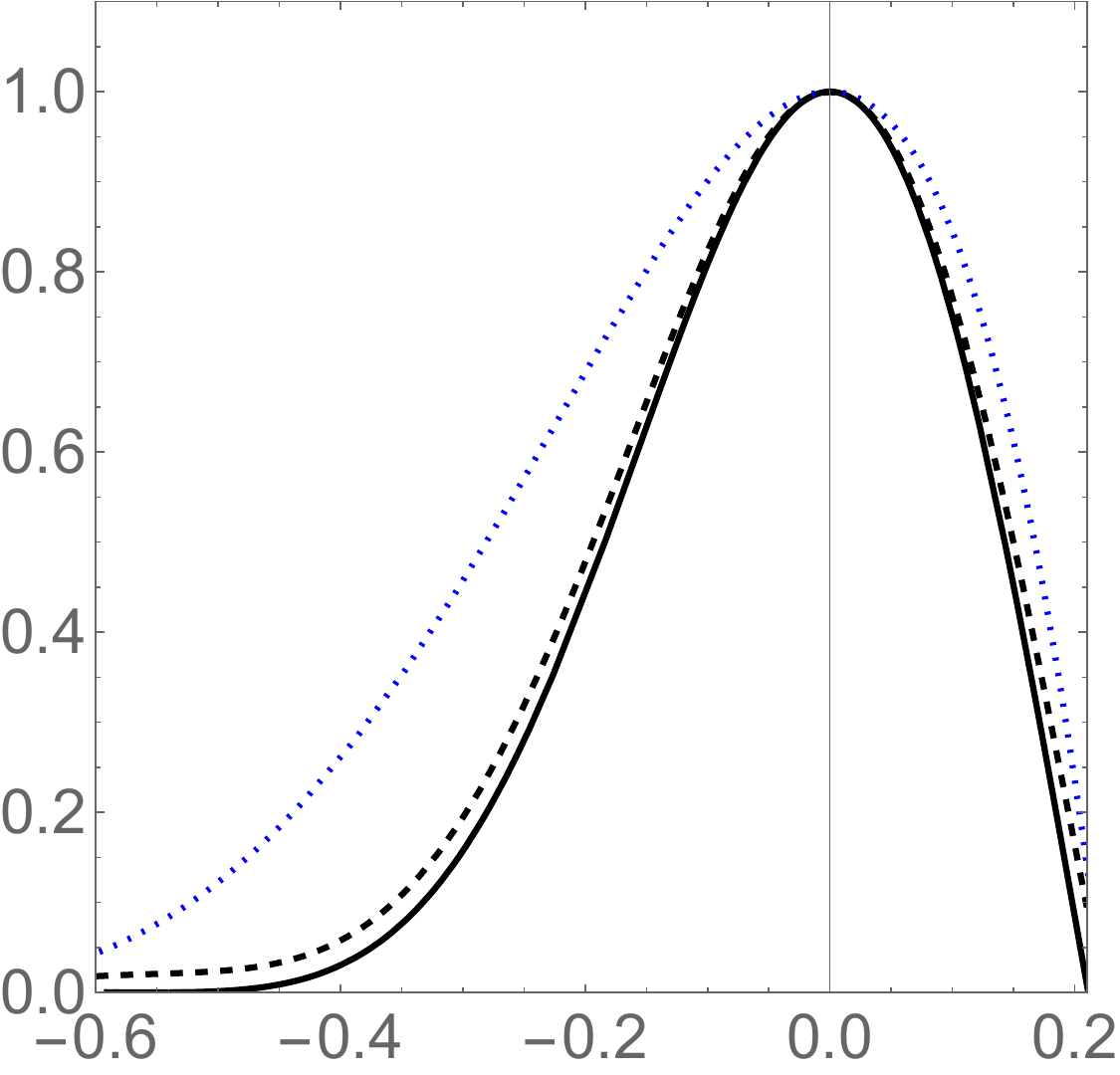}
        \includegraphics[width=0.75\linewidth]{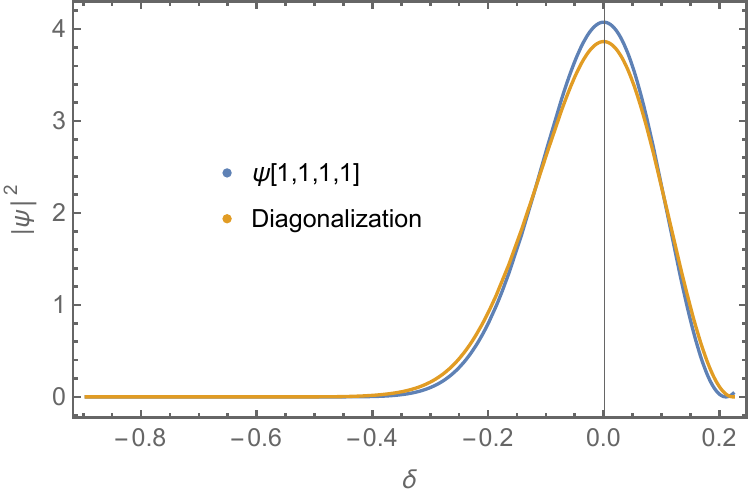}
    \caption{
Upper:    The lowest Laplacian eigenfunction $\psi[1,1,1,1]$ at $\alpha=\beta=\gamma=0$ versus $\delta$ (solid line) compared to our simplest  variational function (\ref{eqn_ansatz_faces}) (dotted line), and its improvement by the Ritz variational procedure (dashed line). For comparison all eigenfunctions are
    normalized to their value at the origin.\\
Lower:  The same  $\psi[1,1,1,1]$ (blue) compared to 
the numerically exact wave function (red) obtained in \protect\cite{PENTAX} and normalized.
    }
    \label{fig_corr1}
\end{figure}

\begin{figure}[b!]
    \centering
    \includegraphics[width=0.75\linewidth]{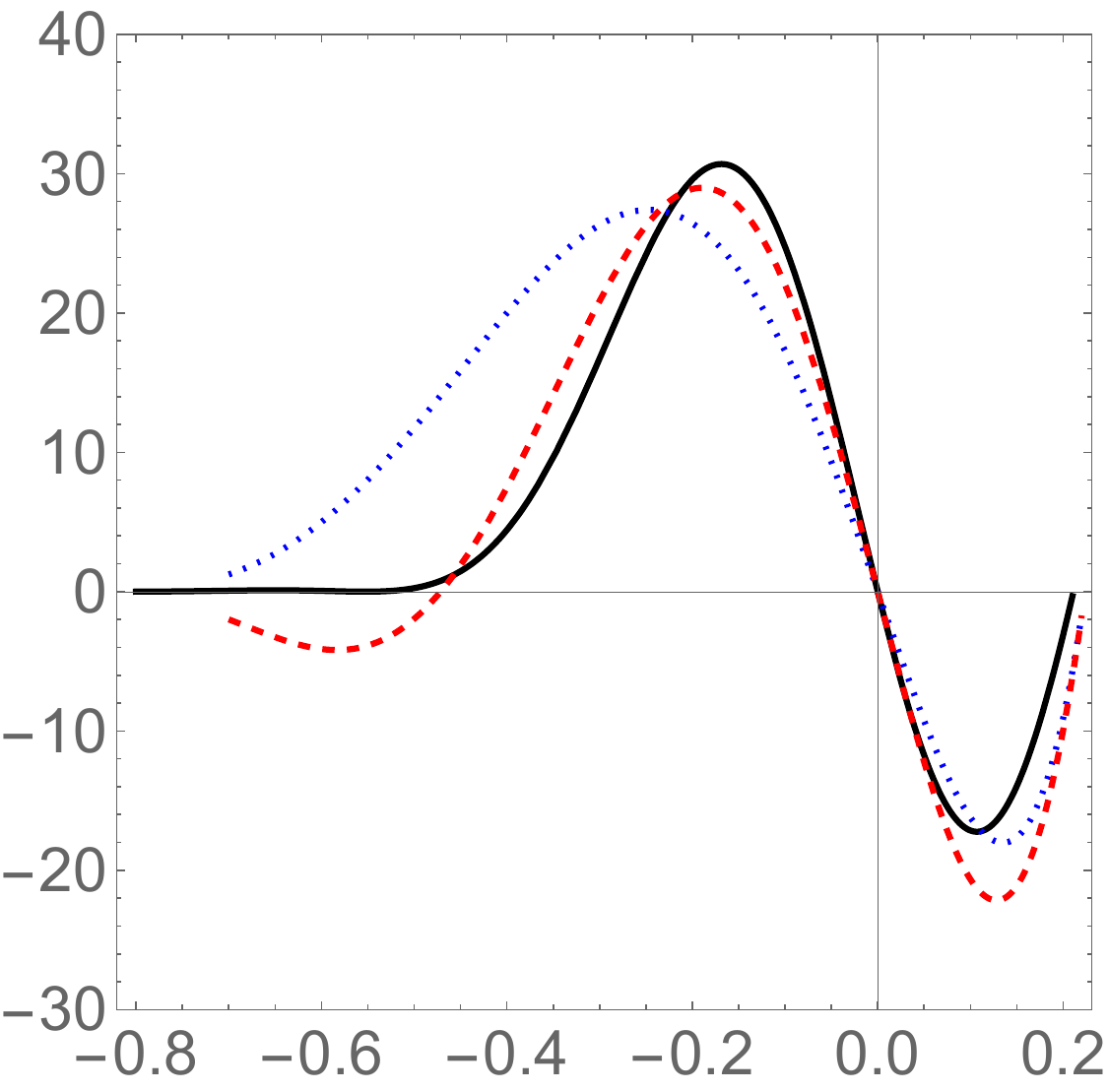}
    \caption{Comparison between the shapes of the P-shell ($L=1$) wave functions (arbitrary units) on the 4-simplex $A_4$, at $\alpha=\beta=\gamma=0$ as a function of $\delta$.
  The black solid line is the exact Laplacian eigenstate with $\psi[\{2,1,1,1\}]$,
   the blue dotted is the initial trial function (\ref{eqn_psi1}) for $C=0$,  and the red dashed lineis the variational function with a single parameter and a value $C=4.5$. }  
    \label{fig_L1_var}
\end{figure}

  In Fig.\ref{fig_corr1} we show the shape of the ground state Laplacian
  eigenfunction on $A_4$, and in Fig.\ref{fig_L1_var} that of the first excited states. While the variational Ritz method proves successful in
  reproducing of the overall shapes of these wave functions (they were
  actually done prior to our discovery of semi-exact eigensystem in the mathematical literature), one also needs to note that they do not 
  accurately describe their tails. 
  

\section{Baryon-pentaquark  mixing on the light front}
\label{sec_mixing}
\subsection{Strategy and definitions}
 In general, the problem of ``dressing" the $qqq$ sector of the baryon WFs
with  extra $\bar q q$ pairs, can be adressed in apparently different ways.  For example, 
in~\cite{Shuryak:2022wtk}  we used a  ``radiative" (DGLAP-like)
description, with $\bar q q$ pair produced from one of the quarks. 
 The mixing was ascribed to the 't Hooft 4-fermion Lagrangian, producing $u\rightarrow u d \bar d $ reaction but $not$ $u\rightarrow u u \bar u $. Another mechanism considered in the litterature, 
is  production via an intermediate pion propagator. The probability, as a function of the momentum fraction of the initial quark,  is convoluted 
  with the PDF of the mother nucleon, e.g. $u^{proton}_v(x)$. In this approach the interference of the 
  newly created quark with those in the nucleon is ignored, in contradiction with Fermi statistics. 
   This may perhaps be justified only if the ``pion-sigma clouds" extend well beyond the original $qqq$ wave function. Another issue with this approach is that the
   calculations are carried at the probability (density matrix, PDF) level.  Our approach is based on the evaluation of the full 5-body wave function as described 
 in our recent study~\cite{Miesch:2025wro}, where one  can find extensive introduction and references.
 
Here we briefly note that the puzzling features of the nucleon PDFs observed experimentally (such as flavor asymmetry of the antiquark sea, or the apparent presence
of significant orbital motion) can find natural explanations
if the extra $\bar q q$ pairs are described as admixture generated by $\sigma$ and $\pi$ mesons. The technical description is detailed in \cite{Miesch:2025wro}, as it follows earlier studies 
where the baryon wave function times pertinent mesonic operators
$T_\sigma,T_\pi$ is folded onto pentaquark states. The so called
meson-baryon-penta overlap is the integral over all coordinates $\vec x_i,i=1..5$. The coordinates allocation is given by 
\be \label{eqn_Cn}
C_n=\int_x\Psi^{penta}_n(\vec x_i) \hat T(\vec x_4-\vec x_5)B(\vec x_1,\vec x_2,\vec x_3) \ee
with the antiquark coordinate identified as $\vec x_5$, and with  all the partonic coordinates  re-expressed in terms of  Jacobi coordinates $\vec \alpha ...\vec \delta$. Each factor, in turn, consists of the ``main" (radial) wave function expressed in  hyperdistances, $Y_5$ for the pentaquark and $Y_3$ for the baryon, times pertinent color-spin-flavor structures (vectors in respective ``monom" spaces). Here $n=1..27$ counts the suitable pentaquark states of the P-shell.

The new  task is to translate these expressions to the  LF formulation. The baryon and pentaquark wave functions are modified as  we now detail.
The nucleon color-spin-flavor structure 
remain as described in previous papers.
The radial wave function $B(Y)$ which depends  on the
hyperdistance $Y_B^2=\vec \alpha^2+\vec \beta^2$, 
splits into the $products$ of two functions depending on the transverse and
longitudinal $momenta$
$$ B(\vec x_i) \rightarrow 
B_\perp(\vec p_i^\perp) B_{\Delta 3}(\alpha,\beta)$$
The latter function $B_{\Delta 3}(\alpha,\beta)$ is the ground state 
baryon wave function in the 2-simplex or the equilateral triangle,
a function of two momentum fractions. 
We recall that for different baryons those are different, e.g. in \cite{Shuryak:2019zhv} we extensively discussed the difference between $\Delta$ and nucleon states, due to a ``good diquark" effect in the latter.

The procedure for pentaquark is similar.
The wave function splits into transverse and longitudinal parts
$$ \Psi(\vec x_i) \rightarrow 
\Psi_\perp(\vec p_i^\perp) \Psi_{\Delta 5}(\alpha,\beta,\gamma,\delta)$$
This splitting reflects the independence of the transverse and 
longitudinal variables in the LF Hamiltonian. The longitudinal wave functions defined in the  4-simplex $\Delta 5,$ was discussed above.

As noted, the nucleon-meson admixture 
must  involves  P-shell pentaquarks, by parity. It is not a small issue, since there is also a part of the WF  in the CM frame which contains $linearly$ dependent coordinates. Since those have a nontrivial transformation matrix under 
quark permutations, it led to extensive 
calculations for orbital-color-spin-flavor 
WFs satisfying Fermi statistics \cite{Miesch:2025wro}. The longitudinal momenta $p_z\gg p_\perp$ follow by boosting to the LF. The component with $L_z=\pm 1$,
built from transverse momenta, is  sub-leading, and only $L_z=0$ survives, with a simplification to $J_z=S_z$.
The longitudinal part is generalized to 
 the orbital-color-spin-flavor WFs found earlier, with the obvious change of the coordinates to longitudinal momentum fractions (also called $\alpha,\beta,\gamma,\delta$). 

The mesonic operators $\hat T$ includes the $\bar q q$
vertex which traditionally is written in coordinates
as Gaussian $$exp[(\vec x_4-\vec x_5)^2/2\rho^2]\,.$$ Its momentum representation is also a Gaussian. We assume that the boost to the LF compresses
hadrons longitudinally, leaving only the factor with
transverse momenta. 

The operators providing proper quantum numbers in $\hat T$ are
$( \vec S \vec L)$ in the $T_\sigma$ and $( \vec S \vec p)$ in the $T_\pi$, respectively. Spin is the
total spin of $q$ and $\bar q$, $\vec S=\vec S_4+\vec S_5$.
The boost leaves rotations only in the transverse plane,
so we use $$( \vec S \vec L)\rightarrow (S_zL_z)$$
In contrast, for the pion-related operator,  the longitudinal
momenta should be dominant, so we use
$$( \vec S \vec p)\rightarrow (S^z_4 p_4+S^z_5 p_5)$$
Obviously, the longitudinal momenta of the 4-th quark and 5-th
antiquark are to be re-expressed in terms of  Jacobi variables $\gamma,\delta$.

In an early  simplified model (with  no projections to pentaquarks) with just the $N\pi,\Delta\pi$ states,
Garvey~\cite{Garvey:2010fi} pointed out that two observables,
the antiquark isospin asymmetry and the orbital momentum, are the same. In our analysis they are not.
In Garvey's model, Fermi statistics is $not$ enforced, 
while in our  approach it is paramount and 
strictly enforced, as all pentaquark states satisfy  
permutation symmetry. 

Finally, the isospins in channels $N+\pi,\Delta+\pi$,
should be analyzed carefully. While 
the total spin $S$,  orbital momentum $L$ and total $J$  do not exist in the LF formulation, the isospin quantum numbers remain unchanged. Adding a sigma meson creates no issues, but adding  a pion, with $I_\pi=1$ to the nucleon or Delta still needs to be done, so that the
 isospin quantum numbers are  the same as that for the proton, namely $I=1/2,I_z=1/2$. This  is done using standard Clebsch-Gordan coefficients
\ba \label{eqn_admixtures}
|N\pi\rangle &=& {1 \over \sqrt{3}} |p \pi^0\rangle-{\sqrt{2 \over 3}}|n \pi^+\rangle \\
|\Delta\pi\rangle &=& {1 \over \sqrt{2}} |\Delta^{++} \pi^-\rangle-{\sqrt{1 \over 3}}| \Delta^+\pi^0\rangle+{1\over \sqrt{6}}| \Delta^-\pi^+\rangle \nonumber
\ea
In general, there are three possible admixtures to the
nucleon, with overlap coefficients called $C_{p\sigma},C_{N\pi},C_{\Delta \pi}$. While at the end their contributions will be added, it is instructive to study 
the three admixtures separately, see Table \ref{tab_admixtures}.

\subsection{The 5-quark sector of the proton}
\begin{figure}
    \centering
    \includegraphics[width=0.95\linewidth]{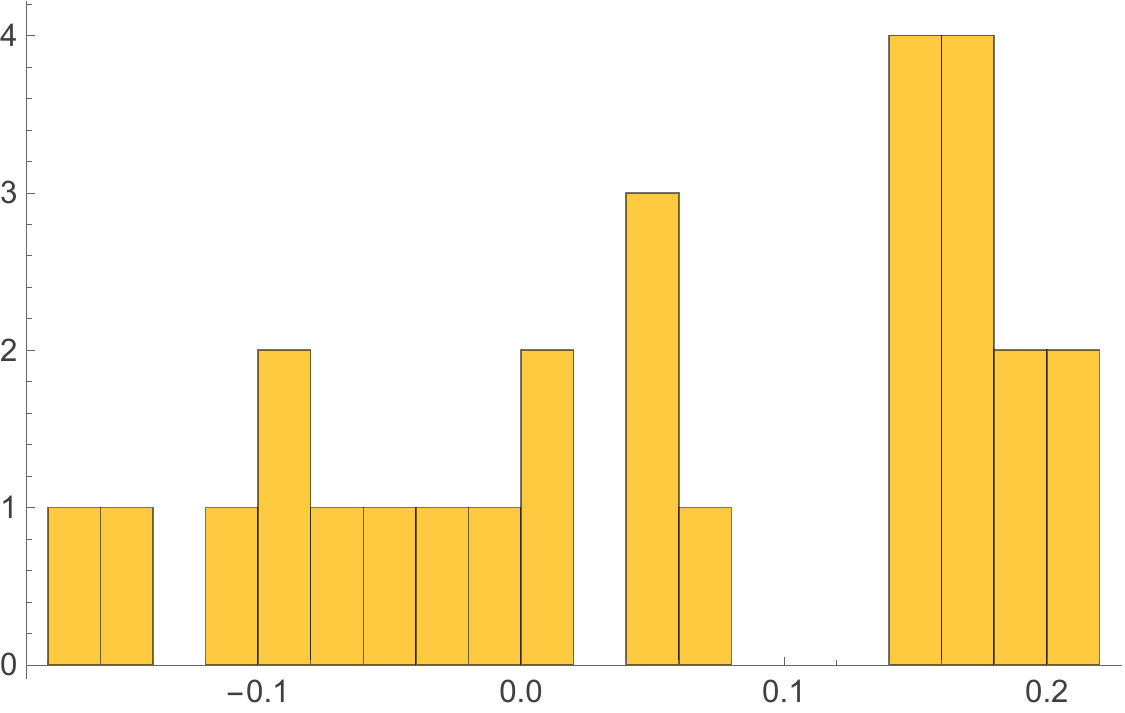}
    \caption{The distribution of the projection
    coefficients $C_n$ of the 27 pentaquark states to
    $N+\sigma$}
    \label{fig_proj}
\end{figure}

Using 27 pentaquark $L=1$ states defined in section \ref{sec_penta}, we have evaluated the projections of
the $p+\sigma,N+\pi,\Delta+\pi$ states onto them. 
The overlap coefficients $C_n$ (\ref{eqn_Cn})
are calculated as explained in the previous section,
using proper combination of coordinates in $|P_n>$,
with the Ansatz function (\ref{eqn_ansatz_faces}).
In this calculation we included only 
the functions normalized in
momentum fractions $\alpha,\beta,\gamma,\delta \in starfish$. Their histogram is given in Fig.\ref{fig_proj}.  It is rather flat, with a maximal modulus around 0.2. There is no dominance of particular states. 

Ignoring the spin-dependent splitting in the denominator
of the first order perturbation theory, we define various
5-quark admixtures to the nucleon as
\be \Delta \psi_{A}= \sum_{n=1}^{27} C_A^n \Psi_n \ee
where the index refers to the three channels $A=p\sigma,N\pi,\Delta\pi$, separately normalize to 1. 
Its structure can be studied both by averaging of 
certain operators, or by integrating over partonic variables.  it has 7200 nonzero
components, in the standard 746496 monom basis. While rather sparse, it is still not small enough to be given
here. 

\begin{figure}
    \centering    \includegraphics[width=0.85\linewidth]{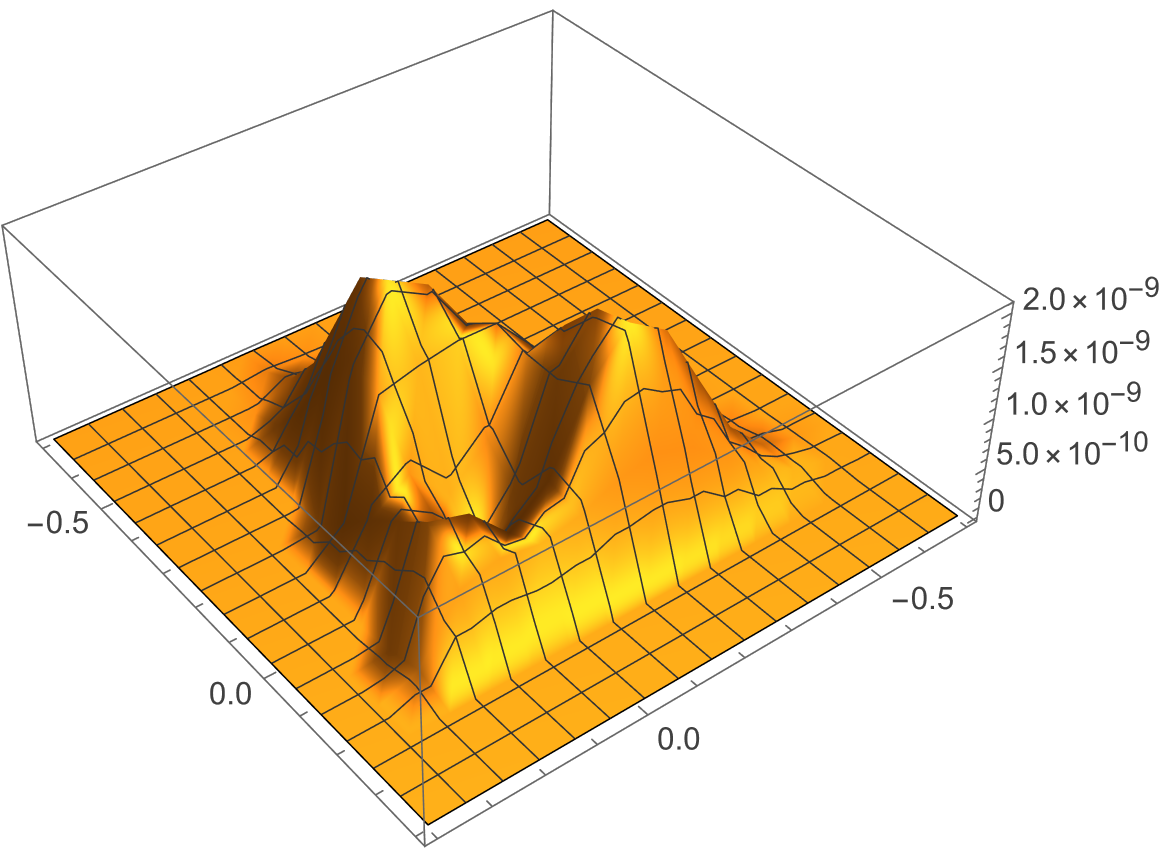}
    \caption{The shape of $\Delta \psi_{N\sigma}^2$ on the $\alpha=\beta=0$ and $\gamma,\delta$ plane}
    \label{fig_Nsigma_shape}
\end{figure}

\begin{figure}
    \centering    \includegraphics[width=0.85\linewidth]{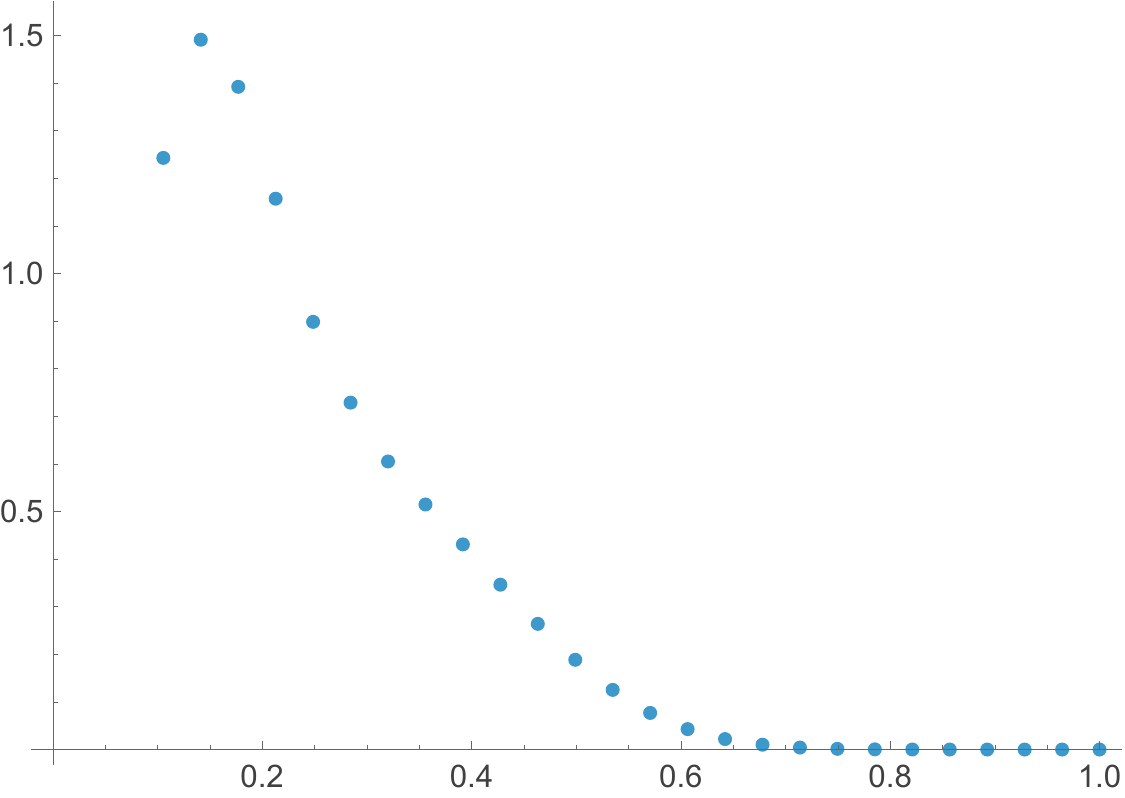}   
    \caption{Predicted PDF of the antiquarks, from $\Delta\psi_{p\sigma}$ (arbitrary units) versus Bjorken momentum fraction $x$. This  plot is linear, emphasising the small x region. Its logarithmic version will be included in Fig.\ref{fig_antiquarks_dglap}.
    }
    \label{fig_Nsigma_PDF}
\end{figure}

The individual 27 pentaquark states are
not spherical, and we project to baryon+meson WF which
designates by 1-2-3 particles in the baryon, and by 4-5 in the mesons. Yet one
of the admixtures turned out to be ``hyperspherical", with a density (squared WF) 
\be \Delta \psi_{p\sigma}.\Delta \psi_{p\sigma} \sim (\alpha^2+\beta^2+ \gamma^2+\delta^2) \ee
However, the appearance of this factor changes the shape of the wave function, so the analogue to Fig.\ref{fig_penta_vary_WF1} now looks  as shown in Fig.\ref{fig_Nsigma_shape}. Furthermore, the $N\pi,\Delta\pi$ projections
generate densities $ \Delta \psi.\Delta \psi$ which are not hyperspherical, with more complicated shapes
\be \Delta \psi_{N\pi}.\Delta \psi_{N\pi} \sim 
 \alpha^2 \gamma^2 +  \beta^2 \gamma^2 + 
  \gamma^4 + 2.68 \delta^4\ee 
  \be \Delta \psi_{\Delta\pi}.\Delta \psi_{\Delta\pi} \sim 
 \alpha^2 \gamma^2 +  \beta^2 \gamma^2 + 
  \gamma^4 + 3.70 \delta^4\ee 
The antiquark PDF
(integral of $\Delta \psi^2$ over $\alpha,\beta,\gamma$)
changes rapidly, see Fig.\ref{fig_Nsigma_PDF}.
Compared to the predictions for the ground
(L=0) shell pentaquarks in Fig.\ref{fig_penta_pdf}, it gets more narrow, and shits  to smaller $x$. Yet the large-$x$ part (better seen in the lower plot) show the
same $\sim (1-x)^8$ trend.

Some properties of three different 5-quark admixtures to the proton  are listed in the table \ref{tab_admixtures}.
Let us start with the mean spin
of the antiquark (particle number 5) 
\be \langle S_5 \rangle =\Delta \psi_{A}. S_5 .\Delta \psi_{p\sigma} \ee
where $A=p\sigma,N\pi,\Delta\pi$. All  show negative contributions, which should be compared   to the total nucleon spin $S_z=+0.5$.

Other important matrix elements refer to the isospin projection of the antiquarks.
Let us define the operators projecting to $\bar d,\bar u$ flavors
\be I_{\pm}= (\hat 1\pm \tau_z)/2 \ee 
applied to particle 5 (antiquark). (Note that the isospin components 
for antiquarks are inverted.) In this case $p\sigma$ and $N\pi$ show
preference for $\bar d$ and the $\Delta\pi$ the other way. 

\begin{table}[]
    \centering
    \begin{tabular}{|c|c|c|c|} \hline
    operator & $p\sigma$ & $N\pi$ & $\Delta \pi$ \\ 
      $S_z(5)$   &  -0.177 & -0.112 & -0.0338 \\
      $\bar d=I_+(5) $ &  0.81  & 0.858 &  0.333\\
           $\bar u =I_-(5) $  &  0.19  & 0.142 &  0.666\\ \hline
    \end{tabular}
    \caption{Some properties of the three 5-quark admixtures to
    the nucleon (\ref{eqn_admixtures}) defined as average values of the
    operators listed in the first column. }
    \label{tab_admixtures}
\end{table}

In summary: the antiquarks in the nucleon are described by baryon-pentaquark mixing, at the WF level. The three contributions to LF  wave function $\Delta \psi_A$, were used to calculate
various matrix elements related to antiquarks, such as
the mean $S_z$ and $L_z$, as well as the mean isospin, and related those with $moments$ of the observed PDFs.

\section{Antiquarks in the nucleon sea: empirical information  versus  theory}
\label{sec_sea}

We now briefly recall some  theoretical ideas 
on the PDFs and similar partonic observables in general. Those are not WFs but
{\em density matrices}, the probabilities to have a given quark/gluon with momentum fraction $x$,  assuming   characteristic resolution scale $Q$, e.g. $x\bar u(x,Q)$.
Compressing a multidimensional wave function to a single-body density matrices is associated
with certain loss of information, thus appearance of the {\em entanglement entropy}. All of that is well known in many applications, e.g. in atomic and nuclear physics. 

The benefit of the LF formulation developed in the previous sections, including the three $\Delta \psi_A, A=p\sigma,N\pi,\Delta\pi$ 5-quark admixtures to the basic 3-q baryons,  is that one can 
 calculate  $any$ partonic observable, e.g. the antiquark PDFs, TMDs, GPds  etc. within the same setting.

Let us also briefly recall, that the  dependence on the momentum transfer $Q$ 
(resolution scale) was developed in QCD first via calculation of the {\em anomalous dimensions} of certain operators using perturbative QCD. That lead to prediction of the Q-dependence of the PDF moments. These moments were later also calculated by lattice QCD
practitioners.

Dokshitzer-Gribov-Lipatov-Altarelli-Parisi (DGLAP)
developed a ``radiative evolution" of PDFs, 
in the form of integro-differential equations. 
In order to compare  experimental data
taken at various resolutions, the results are usually ``evolved" to some standard scales. Large groups of people and plenty of papers were done, using ``global PDF fits", with increasing accuracy. They are numerous to refer to here. .
The tutorials/references are given e.g. at the web page of the CTEQ collaboration. Here, we  briefly illustrate some  features using the ManeParse Mathematica package. Those are based on fits to whole volume of data, for deep inelastic scattering and Drell-Yan processes, which are mostly sensitive to antiquark PDFs.

 As we discussed in our previous papers, the theory we tried to develop aims at including the ``instanton-based chiral physics", which is   dominant at the resolution of $Q^2\sim 1 \, GeV^2$, or of the order of the instanton size $\rho\sim 1.4 \, GeV^{-1}$. To match
to  partonic observables, the latters should be
 evolved downward, to this ``low scale". We now proceed
to directly interpolate the empirical PDFs to this low scale.

\begin{widetext}

\begin{figure}[t!]
    \centering
    \includegraphics[width=0.45\linewidth]{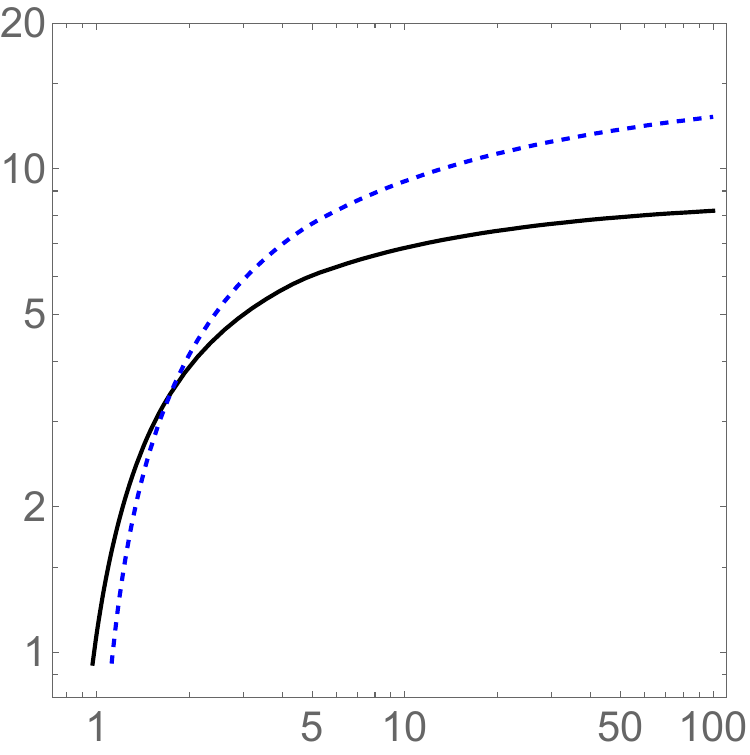}  
    \includegraphics[width=0.45\linewidth]{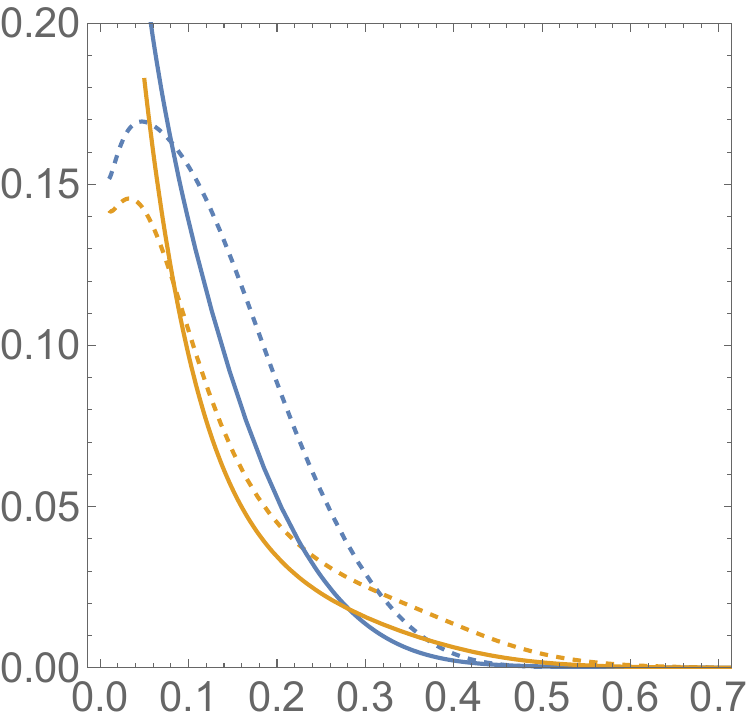}
    \caption{Left: Evolution of the gluon PDF $xg(x,Q)$
    versus $Q$, at $x=0.01$ (black solid) and  $x=0.005$ (blue dashed). It shows the disappearance of the soft gluons at
    $Q\rightarrow 1\, GeV$.\\
  Right:  The antiquark $x\bar d(x,Q),x\bar u(x,Q)$ distributions (blue and red) in $x$,
  at the resolution $Q^2=50\, GeV^2$  (solid lines) evolving down 
  to low resolution $Q^2\sim 1\, GeV^2$ (dashed lines).\\ Both plots were generated
  by ManeParse Mathematica package for generating PDFS, arXiv 1605.08012. It includes DGLAP evolution and
  interpolation of multiple data sets done by the CTEQ collaboration.
    }
    \label{fig_antiquarks_dglap}
\end{figure}
\end{widetext}

 In Fig.\ref{fig_antiquarks_dglap} we first show examples of the DGLAP
 evolution of the $gluon$ PDF, from the highest $Q$ on  the r.h.s. to the resolution of  $Q=1\, GeV$.
 Note that it is in log-log scale. The chief observation is that multiple soft gluons are present at the  experimental resolution, that totally  disappear at the scale $Q=1\, GeV$. This scale is dominated by the semi-classical glue, not the perturbative gluons.

 The right plot shows the empirical antiquark PDFs. The solid lines are  taken at the scale $Q^2\sim 50 \, GeV^2$, more or less the scale in Drell-Yan processes. 
In order to  be compared with
our theory (keeping only quarks and antiquarks), they
 need to be evolved down to the  
scale $Q^2\sim 1 \, GeV^2$ where gluons disappear.
The evolved results are illustrated  by the dashed curves.
%

 A clear feature is
the different $\bar u$ and $\bar d$ curves, or ``flavor asymmetry to be discussed below.
The shape of the distributions changes, as they move to the right (larger $x$), as shown in Fig.\ref{fig_antiquarks_dglap} (right).

\begin{figure}
    \centering
    \includegraphics[width=0.85\linewidth]{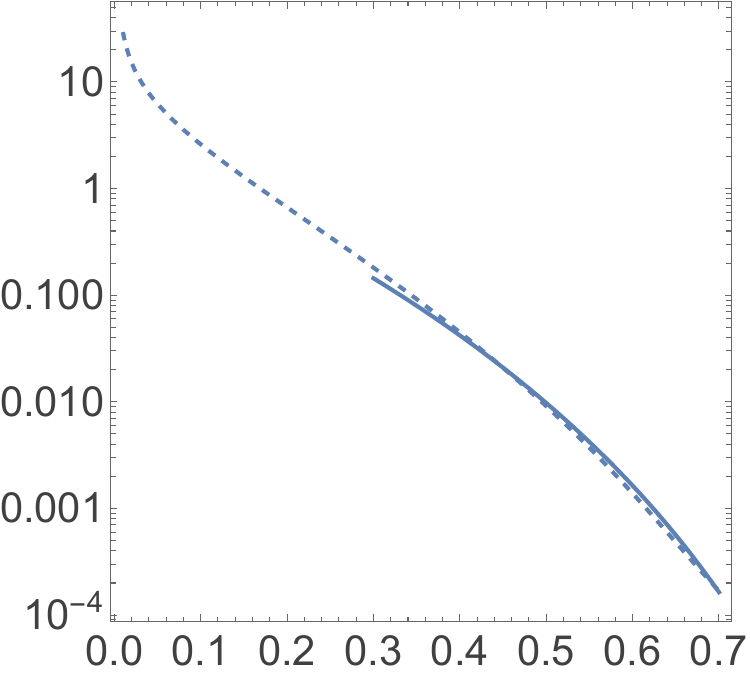}
    \caption{The PDFs of all the antiquarks $u(x,1)+d(x,1)$ at the low scale $Q=1\, GeV$ (no $x$ prefactor here), from ManeParse, compared to the fit $\sim (1-x)^8$ 
    shown by the solid line.\\
   } \label{fig_antiquark_PDF_Q1}
\end{figure}

Another feature is the very rapid decrease towards  larger $x$. This feature is in qualitative agreement 
with our pentaquark admixed WFs.  To make this more quantitative, we  also show the empirical (fit) to the antiquark PDF in Fig.\ref{fig_antiquark_PDF_Q1} (logarithmic scale), emphasizing the large-$x$ tail  $\sim (1-x)^8$. In reality, we do not think that the available experimental data  on the antiquarks with $x>1/2$, are reliable enough for an accurate estimate of the tail.
 
Our analysis based on a Hamiltonian with only kinetic and confinement (no spin effects), but with full 5-quark admixtures and enforced Fermi statistics, suggests  an anti-quark PDF with a hard edge  $\sim (1-x)^{12}$. The robustness of this prediction maybe subject to changes
when other effects are included, a subject for future investigations.

\section{Summary and Discussion} \label{sec_summary}
This paper follows our previous work \cite{Miesch:2025wro}, in which we worked out
the WFs  and spectra of light pentaquarks, as well as their role in
explaining the 5-quark sector of the nucleon.
$$ | N >= C_3 |3q >+ \sum_i C^A_5 |5q,A >+... $$
The nontrivial aspects of the calculation 
was the requirement that the admixed pentaquarks must belong to their P-shell with $L=1$,
thus leading to a complex orbital-color-spin-flavor wave function. It was obtained
by fixing the antisymmetric representation
of the permutation group $S_4$, as required by Fermi statistics.

Our earlier work 
was done in the CM frame, and in this work
we generalized it to the light front. Pentaquark wave function have longitudinal
WFs valued in 4-simplex (``starfish"), in a 4d 
space of Jacobi momentum fractions. We obtained
this wave functions both variationally and using an exact
mathematical procedure, including a sum over 
many sinusoidal waves.

The 5-quark sector is the first one with  antiquarks
in the nucleon wave function. The chieff goal was
to extract their perinent  PDFs, and compare them to available experimental data. The antiquark PDF for the ground S-shell
pentaquark were shown in Fig \ref{fig_antiquark_PDF_Q1},
those for 5-quark admixture to the nucleon in Fig.\ref{fig_Nsigma_PDF}. The experimental antiquark PDFs  evolved to the low resolution point $Q^2=1\, \rm GeV^2$ were given in Fig\ref{fig_antiquarks_dglap},
in  qualitative agreement with our results.

{\centerline{\bf Acknowledgements}}
\vskip 0.5cm
This work is supported by the Office of Science, U.S. Department of Energy under Contract  No. DE-FG-88ER40388.
This research is also supported in part within the framework of the Quark-Gluon Tomography (QGT) Topical Collaboration, under contract no. DE-SC0023646.

\appendix

\section{Baryons and tetraquarks on the light front} \label{sec_bar}
Our investigation of multi-quark hadrons on the LF, started with  $qqq$ baryons and their WFs~\cite{Shuryak:2022thi}. In this Appendix we briefly recall some of the facts and notations for completeness.

At the start of \cite{Shuryak:2022thi} we addressed
the issue of perturbative and nonperturbative interquark forces in baryons. While in the literature these forces are mostly discussed using  {\em Ansatz Y} (three strings joined by a color junction), both lattice-based studies as well as our own instanton-based results suggest that the so called {\em Ansatz A} seems to be preferred, for which a ``half-perimeter" appears.
We used the einbein trick
\ba  H_{confinement} &=& (\sigma/2) \big[ r_{12}+r_{13} + r_{23}\big] \rightarrow \nonumber \\
&=& (\sigma/4) \big[{3 \over a}+ a(r_{12}^2+r_{13}^2 + r_{23}^2)\big] \rightarrow \nonumber \\
 &=& (\sigma/4) \big[{3 \over a}+ 3a(r_\rho^2+r_\lambda^2)\big]
\ea
and factored the CM out by using Jacobi coordinates, e.g.  $\vec r_\rho=i\partial/\partial \vec p_\rho$ in recast in the momentum representation ($\hbar=1$). For Ansatz Y the same expression appears, with a different tension $\sigma_A\approx  0.87\,\sigma_Y $.
The longitudinal part of the kinetic energy takes the part of Laplacian on the triangle
\be H_{triangle} =-{\sigma_Y \over a}[{\partial^2 \over \partial \rho^2}+{\partial^2 \over \partial \lambda^2}]\ee
and the transverse part of the Hamiltonian
\be H_{\perp}=3(\vec p_\rho^2+\vec p_\lambda^2)-3m_Q^2({\sigma_Y \over a})({\partial^2 \over \partial \vec p_\rho^2} +{\partial^2 \over \partial \vec p_\lambda^2 }) \ee
is that for harmonic oscillator.
The eibein parameter $a$ is found by 
minimization, as illustrated in Fig.~8 in \cite{Shuryak:2022thi}. Its value depends on the quark mass in question. 

 The only non-factorizable part of the problem -- which makes it challenging -- is the  ``effective potential"  depending on both the transverse and longitudinal variables
\be V_{cup}  \sim \sum_{i=1}^3 (\langle \vec p_{i,\perp}^2 \rangle +m_i^2) \big[{1 \over x_1}+{1 \over x_2}+{1 \over x_3}-9 \big]  \ee
The approximation is in changing the transverse momenta squared to their average value.
We then subtract a constant so that at the center of the triangle the parton momenta $x-1=x_2=x_3=1/3$
are null. 

In~\cite{Shuryak:2022thi} the ``cup potential" $V$ 
was recast in  matrix form $\langle n | V | n'\rangle $,
with a finite cut-off in the basis set. However, the  sensitivity to this cutoff was not investigated. 
Alternatively,  the Schrodinger equation on a triangle with Dirichle boundary conditions can be carried numerically, using in Mathematica the $NDEigensystem$ command, with the results shown in Fig~\ref{fig_bar_wf}.

For three bodies, $N=3$, all eigenstates of the Laplacian free of the CM of motion 
were derived analytically in~\cite{Shuryak:2022thi}.
We now report further progress in this direction. These states can be   written in the
form of a Slater determinant 
\begin{widetext}

\begin{figure}[t!]
    \centering
    \includegraphics[width=0.45\linewidth]{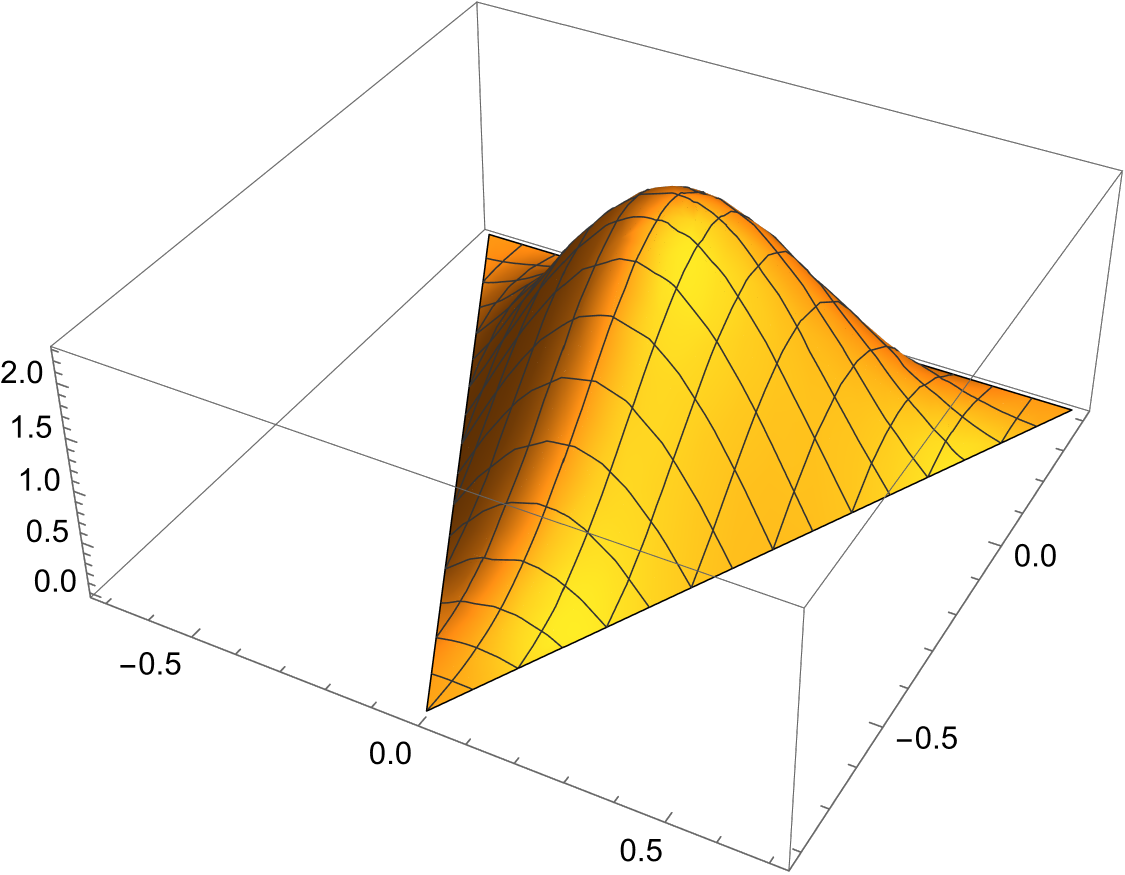}
        \includegraphics[width=0.45\linewidth]{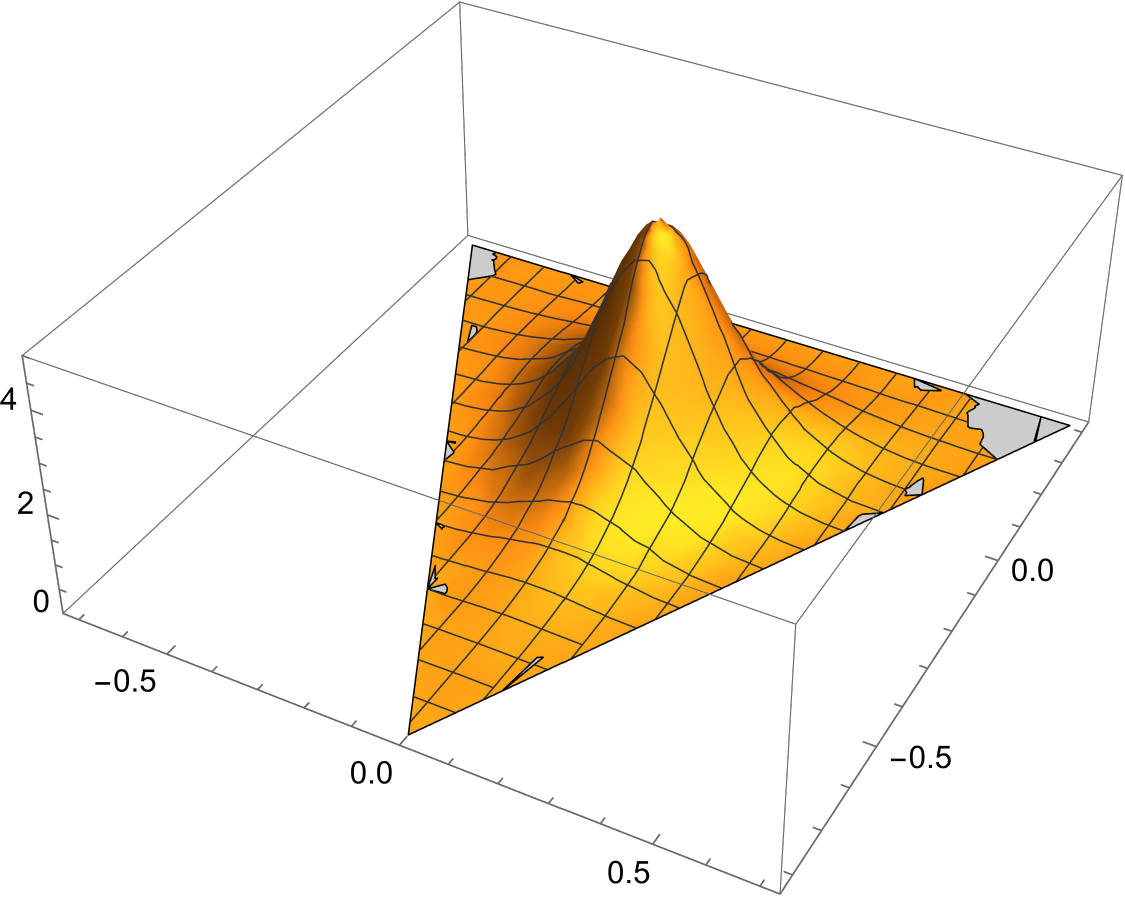}
    \caption{Forward part of baryon wave function on the LF. The left  plot is for only the Laplacian, and the right plot for the  Laplacian plus the cup potential.}
    \label{fig_bar_wf}
\end{figure}

\bea
\label{DET3}
\varphi_n[\tilde s]=
\frac{1}{(2\pi)^{3/2}\sqrt{3!}}\begin{vmatrix}
   1& 1   & 1\\
    e^{i\tilde  n_{21}\tilde s_1}& e^{i\tilde  n_{21}\tilde s_2}  & e^{i\tilde  n_{21}\tilde s_3} \\
      e^{i\tilde  n_{31}\tilde s_1}& e^{i\tilde  n_{31}\tilde s_2}  & e^{i\tilde  n_{31}\tilde s_3} 
\end{vmatrix}
\eea
with the  spectrum
\bea
\tilde E[\tilde n]=\frac {16\pi^2}3(
 n^2_{21} +  n^2_{31}
-  n_{21}  n_{31})
\eea
The spectrum in our work
\bea
e^D_{m_L,n_L}=\frac{16\pi^2}{9L^2}(m_L^2+n_L^2-m_Ln_L)\nonumber\\
\eea
with $L=\sqrt 2$ coinsides with it.  The normal coordinates  can be identified with the standard Jacobi $\rho=\alpha, \lambda=\beta$ coordinates used in \cite{Shuryak:2022thi}
\bea
s_1-\frac SN&=&
\frac 1{\sqrt 6}(\lambda+\sqrt 3\rho)\nonumber\\
s_2-\frac SN&=&
\frac 1{\sqrt 6}(\lambda-\sqrt 3\rho)\nonumber\\
s_3-\frac SN&=&
-\sqrt{\frac 23}\lambda
\eea
If we set $\tilde n_{21}=2\pi n_L$ and
$\tilde n_{31}=2\pi (2n_L)$ to compare with the non-degenerate Dirichlet states
reported in~\cite{Shuryak:2022thi}, then the determinantal wavefunction reduces
to
\bea
\varphi_{n_L}[\rho, \lambda]=\frac{2^3(-1)^{n_L+\frac 12}}{(2\pi)^{\frac 32}\sqrt{3!}}\,
 {\rm sin}\bigg(\frac{2\pi n_L\rho}{L}\bigg)
{\rm cos}\bigg(\frac {\pi n_L}L(\rho+\sqrt 3\lambda)\bigg)
{\rm cos}\bigg(\frac {\pi n_L}L(\rho-\sqrt 3\lambda)\bigg)
\eea
For the ground state with $n_L=1$, it can be recast in terms of 3 standing sine waves after a shift in $\rho=\tilde\rho+L/2$, which define the boundary or 3 edges  of the 2-simplex (triangle)
\bea
\varphi_1[\rho,\lambda]\rightarrow
\frac {2^3}{(2\pi)^{\frac 32}\sqrt{3!}}
\,{\rm sin}\bigg(\frac{2\pi \tilde\rho}L\bigg)
{\rm sin}\bigg(\frac{\pi}L(\tilde\rho+ \sqrt 3\lambda)\bigg){\rm sin}\bigg(\frac{\pi}L(\tilde\rho- \sqrt 3\lambda)\bigg)
\eea

The ground state WF  for the $N=4$ Laplacian
is again a product of 6 standing sine functions, 
\bea
\varphi_1[\rho, \lambda, \xi]=&&
\frac {2^4}{\sqrt{(2\pi)^4 4!}}\,
{\rm sin}\bigg(\frac {2\pi \rho}L\bigg)
{\rm sin}\bigg(\frac {2\pi \lambda}L\bigg)
\nonumber\\
&&\times
{\rm sin}\bigg(\frac {\pi}L
(\lambda+\rho+\sqrt 2\xi)\bigg)
{\rm sin}\bigg(\frac {\pi}L
(\lambda+\rho- \sqrt 2\xi)\bigg)
{\rm sin}\bigg(\frac {\pi}L
(\lambda-\rho+ \sqrt 2\xi)\bigg)
{\rm sin}\bigg(\frac {\pi}L
(\lambda-\rho- \sqrt 2\xi)\bigg)\nonumber\\
\eea
\end{widetext}
For $L=1$, the waves vanish at the edges of the regular tetrahedron delimited by the 4 planes
\bea
\lambda+\rho+\sqrt 2\xi=1\nonumber\\
\lambda-\rho+\sqrt 2\xi=1\nonumber\\
\lambda+\rho-\sqrt 2\xi=1\nonumber\\
-\lambda+\rho+\sqrt 2\xi=1
\eea

Alternatively, for the regular tetrahedron 
embedded in the 3-cube with coordinates $x,y,z\in [0,1]$ 
and  defined by the 4 planar face conditions
\bea
1-x+y-z\geq 0\nonumber\\
1-x-y+z\geq 0\nonumber\\
1+x-y-z\geq 0\nonumber\\
x+y+z-1\geq 0
\eea
we may identify  the coordinates $s_i$
in the Fermion determinant, with the  distances to the 4 planar faces of the tetrahedron
of edge lenght $\sqrt 2$, 
\bea
\label{LIN5}
{s_1}&=&\frac 1{\sqrt 3}(1-x+y-z) \nonumber\\
{s_2}&=&\frac 1{\sqrt 3}(1-x-y+z) \nonumber\\
{s_3}&=&\frac 1{\sqrt 3}(1+x-y-z) \nonumber\\
{s_4}&=&\frac 1{\sqrt 3}(x+y+z-1) 
\eea
 with constant $$\sum_{i=1}^4 s_i=S= {2\over \sqrt 3}$$ by Fermat theorem. The cartesian variables $x,y,z$ are related linearly to the Jacobi variables $\rho, \lambda, \xi=\alpha, \beta, \gamma$. In terms of (\ref{LIN5}), the free Laplacian is no longer diagonal in the new $x_i=x,y,z$ and $S$  coordinates, reducing to the general elliptic operator
\bea
\sum_{i=1}^4\partial^2_{s_i}=&&\frac 14(
7\partial^2_x+7\partial^2_y+7\partial^2_z+13\partial^2_S\nonumber\\
&&8\partial^2_{xy}+8\partial^2_{xz}+8\partial^2_{yz}
\nonumber\\
&&+14\partial^2_{xS}+14\partial^2_{yS}+14\partial^2_{zS})\nonumber\\
\eea
Note that the stationary WFs
\bea
\label{SATZ1}
&&\varphi_{n_1n_2n_3n_4}(x,y,z)=\nonumber\\
&&{\rm sin}\bigg(\frac{n_1\pi s_1}S\bigg){\rm sin}\bigg(\frac{n_2\pi s_2}S\bigg){\rm sin}\bigg(\frac{n_3\pi s_3}S\bigg){\rm sin}\bigg(\frac{n_4\pi s_4}S\bigg)\nonumber\\
\eea
diagonalize the 4-dimensional free Laplacian (CM unsubtracted) with energies
\bea
e_{n_1n_2n_3n_4}=\sum_{i=1}^4\bigg(\frac{n_i\pi}S\bigg)^2
\eea
and vanish identically on the 4 faces of the regular tetrahedron. The generalization of this  approach to 5-bodies
is  much more involved, a subject of a separate
analysis in~\cite{PENTAX}. We now will turn to a simpler
variational approach.

\section{Variational calculations of the longitudinal part of the pentaquark states}
\label{sec_var}

We start by recalling that for five bodies, $N=5$, the kinematical domain is that of a regular $pentachoron$, a 4d simplex with 5 corners, 
10 edges, 10 equilateral faces and 5 regular tetrahedrons, we also referred to as  ``a starfish". There are basically two ways to solve Schrodinger equation
in this geometry: (i) use a variational Ritz method, to be followed in this work,
and (ii) use a full basis set for a different pentachoron (appearing
in the problem of $N$ fermions on the circle) which we presented in~\cite{PENTAX}.

\begin{figure}
    \centering
    \includegraphics[width=0.75\linewidth]{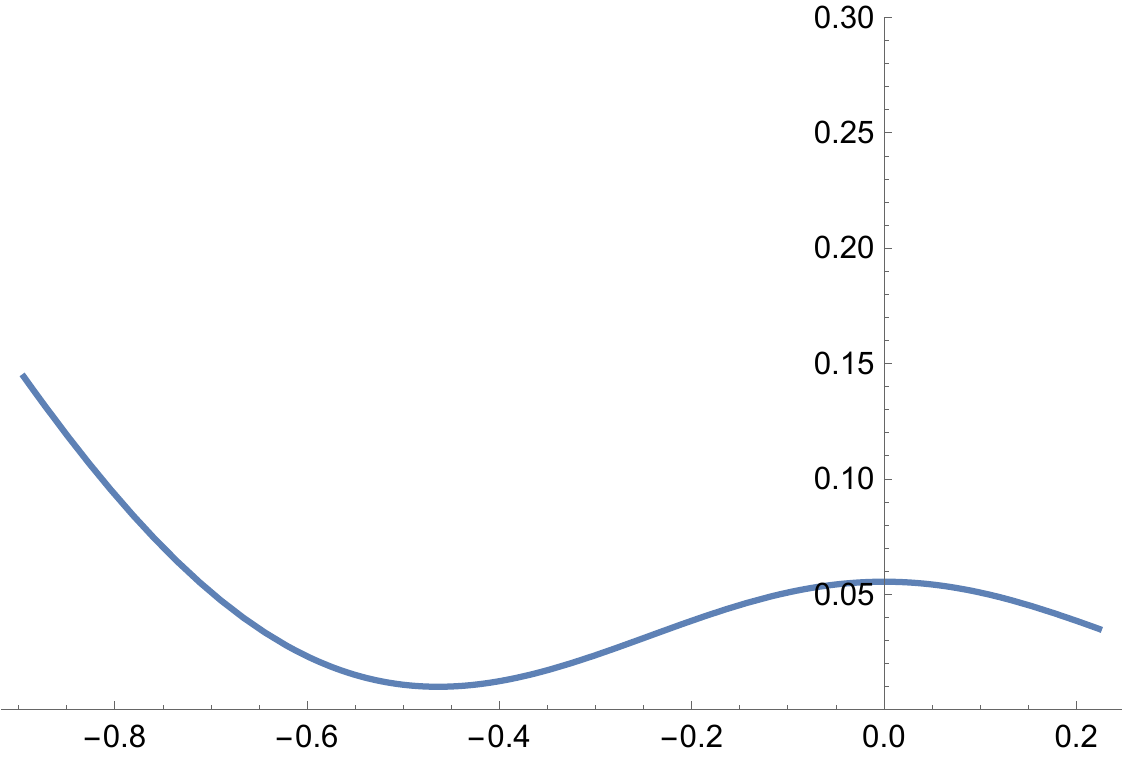}
    \caption{The sum of Gaussians in the correction function (\ref{eqn_Ritz}) after minimization of the Laplacian/norm ratio, N=3.
    The plot is for  $\alpha=\beta=\gamma=0$ as a function of the last Jacobi momentum fraction $\delta$. The left side is the ``corner" and the right side the ``face" of the simplex, where the Ansatz itself vanishes. }
    \label{fig_corr_s}
\end{figure}

The solution of the elliptic PDEs by the Ritz method is well known. It consists of minimizing the corresponding  energy functional over a set of parameters, usually coefficients of certain basis functions. For the ground state, we found a relatively simple set of Gaussians trial functions
\be \label{eqn_Ritz}
\Psi_{Ritz}= \Psi_{\rm faces}*\bigg(1+\sum_i^N C_i exp(-Y^2/2R_i^2)\bigg)\ee
where the first term is the linear Ansatz (\ref{eqn_ansatz_faces}),
and the bracket gives a ``correction function". The average value of the Laplacian (with and without the cup potential) divided by a normalization
$$ \langle - \nabla^2 \rangle ={\int_{simplex}\Psi_p(- \nabla^2) \Psi_p  \over \int_{simplex} \Psi_p^2} $$
 is minimized over $C_i$ (using the $NMinimize$ command). We used various number of correction terms and different sets of radii, eventually returning to just three Gaussians. For a Laplacian/norm we found the improvement, from the initial value of $182.$ to $165$,
 improvement
 by about 10\%, with a differential correction shown in Fig.\ref{fig_corr_s}.
The orrections are  relatively small, at the scale of few percents.  

We found that adding  
the cup potential, $ V_{cup}\sim (1/x_1+ 1/x_2+1/x_3+1/x_4+1/x_5-25)$
amounts  to a  $decrease$ of the correction function, making the
ground states WF close to the original linear Ansatz (\ref{eqn_ansatz_faces}). So, we have used it in the calculations to follow.

The P-shell pentaquark states have negative parity, due to the  multiplication by the extra linea coordinates. The simplest trial function that minimizes the Laplacian  is of the form
\be \label{eqn_psi1}
\psi_1(C)\sim \Psi_{faces}\delta \big[1-C(\alpha^2+\beta^2+\gamma^2+\delta^2) \big]  \ee
The unmodified function $\psi_1(0)$ gives 
$\langle - \nabla^2 \rangle / \langle 1\rangle\approx 312.0 $
The modified one with $C=4.5$ reduces it  by about 1/10. More
complicated trial functions we have tested lead to some improvements. The actual shape of the functions are compared in Fig.\ref{fig_L1_var}.  (Note that the change of sign near zero is a robust
feature, while another zero near $\delta\approx -0.5$ in the variational result is not as robust and is ansatz dependent. Clearly,
the second zero should not be present in the true P1 state. (Yet we checked that this function
is orthogonal to the S-shell $L=0$ variational function, within small integration errors.) We conclude that the variational method is not
to be trusted for $\delta<-0.5$.

\section{Exact  Laplacian eigenstates }
The ``equilateral" (or regular) 4-simplex on which we need to solve the
LF Schrodinger equation is known in mathematical literature as the $A_4$
simplex. It is related to the representations of the permutation group $S_4$.
The theory of tiles in arbitrary dimensions are related with diagonalization of the Laplacian, that we only learned of when most of the present paper was completed.

For the presentation, we follow the derivation in \cite{Doumerc_2008}. The so called {\em fundamental weights},  in $dim=5$ are four
vectors $\omega_i$ in the 5d space
\ba &\omega&=\{ \{ 4/5, -1/5, -1/5, -1/5, -1/5\}, \nonumber \\
&\{&3/5, 3/5, -2/5, -2/5, -2/5\}, \nonumber \\
&\{&2/5, 2/5, 2/5, -3/5, -3/5\}, \\
&\{&1/5,  1/5, 1/5, 1/5, -4/5\} \}\nonumber
\ea
Together with the origin  $\{0,0,0,0,0\}$, these vectors  make 5 corners of a simplex,  which is similar but different from the ``starfish" simplex we need. 
In Jacobi coordinates its corners lie at 5 different  points $a_i$ 
\ba &\{& 1/\sqrt{2}, 1/\sqrt{6}, 1/(2 \sqrt{3}, 1/(2 \sqrt{5})\}, \nonumber \\
 &\{&-(1/\sqrt{2}), 1/\sqrt{6}, 1/(2 \sqrt{3}), 1/(2 \sqrt{5})\}, \nonumber \\
&\{&0, -\sqrt{2/3}, 1/(2 \sqrt{3}), 1/(2 \sqrt{5})\}, \nonumber \\
&\{&0, 0, -(\sqrt{3}/2), 1/(2 \sqrt{5})\}, \nonumber \\
&\{& 0, 0, 0, -(2/\sqrt{5})\}
\ea
\begin{widetext}
 The affine transformation from one simplex to the other
can be  found numerically, using the Mathematica command {\em FindGeometricTransform}. It follows from rotation and shifts. 
\ba &\{& \{0.4 + 0.228825 \alpha + 0.213762 \beta + 0.237755 \gamma + 
  0.473607 \delta \}, \nonumber \\
 &\{&0.2 - 0.59907 \alpha - 0.0816497 \beta + 0.0288675 \gamma + 
  0.111803 \delta, \}\nonumber \\
&\{& + 0.228825 \alpha - 0.559635 \beta - 
  0.11547 \gamma, \} \nonumber \\
 &\{&  -0.2 + 0.0707107 \alpha + 0.305049 \beta - 
  0.518007 \gamma - 0.111803 \delta, \} \\
 &\{& -0.4 + 0.0707107 \alpha + 
  0.122474 \beta + 0.366854 \gamma - 0.473607 \delta \} \} \nonumber\ea

In any dimension,  the eigenstates of the Laplacian can be written as a sum
over permutations over the whole $S_{dim-1}$ group. 
In Mathematica it has a compact form
\be
Sum[Signature[w] Exp[2 I*Pi*(Permute[p, w] . y)], \{w, Sn\}]]]
\ee
or
\be
    \sum_{w\in S_{d}}\text{sign}[w] e^{2\pi i (wp).y}
\ee
The ``momenta" $p=\sum_1^4 m_i\omega_i$ are defined with $m$ ranging over a 4-dimensional set of positive integers, $Sn$ being all the $(d-1)!$ permutations of the permutation group. Here $y$ are the coordinates, either in the original simplex, or those projected to the ``starfish"   as explained above.  The permutations of the components of $p$ according to $w$ (not $\omega$!) are necessary to account for every edge of the simplex.
\end{widetext}

The ground state correspond to $m=\{ 1,1,1,1\}$ or $p=\{-2,-1,0,1,2\}$,  a set of
momenta  in the ``Fermi sphere" with 5 elements.
The exact form of its real part is given at the end of the section due to its complexity. The imaginary part of the sum is identically zero, so we only quoted the real part.

These set of functions  are eigensstates of the Laplacian. The lowest eigenvalue of $\psi[1,1,1,1]$ is 
\be 
-\nabla^2\psi[1,1,1,1]/\psi[1,1,1,1]= 186.972\ee
which is in the ballpark of the values found using the  variational minimization. Recall that the numerically
exact lowest eigenstate yields a lowest eigenvalue of 164~\cite{PENTAX}.
The shape of $\psi[1,1,1,1]$ is compared to the variational and numerically exact solution in Fig.~\ref{fig_corr1}. The $\psi[1,1,1,1]$ eigenfunction is concentrated near the origin, and is smaller near the corner (l.h.s. of the plot).

However, we note that the two set of functions exhibit differences.  The set $\psi[m_1,m_2,m_3,m_4]$
enforces the Laplacian  at {\em any point} inside the 4-simplex, but only enforces the boundary conditions  on a subset of the boundary conditions (e.g. corners and edges of the 4-simplex). In contrast, all  variational functions are designed to vanish on the whole boundary,
but with varying Laplacian in bulk. (Recall that the minimization in the previous Appendix,
was defined as a ratio of $integrals$ over the simplex, as suggested by the Ritz method.) 

To quantify the deviations of $\psi[1,1,1,1]$ from the strict Dirichlet conditions, let us compare its values at the boundary to that at the center.  Here, and for clarity, the arguments are the set of 5 Bjorken momentum fractions, with the normalization set by the value of the WF at the center  of the 4-simplex or  $\psi_0=\psi\{1/5,1/5,1/5,1/5,1/5\}$
\ba
&\psi&\{1,0,0,0,0\}/\psi_0= -6.75737*10^{-6} \, \,\,(corner) \nonumber \\
&\psi&\{1/2,1/2,0,0,0\}/  \psi_0=0.000133 \, \,\,(edge)  \\
&\psi&\{1/3,1/3,1/3,0,0\}/  \psi_0=0.007927 \,\, (3face) \nonumber \\
&\psi&\{1/4,1/4,1/4,1/4,0\}/   \psi_0=0.094441 \,\, (4face) \nonumber
\ea
The first value at the corners is consistent with
the expected numerical accuracy, while the others are not. We also note that the 5-fold symmetry is holding strictly, e.g. the values at all corners are the same.

The nonzero value of the function at the boundary leads to a logarithmic divergence of the mean value of the potential 
 $\int_{\rm simplex} \psi^2 V$,  which in the infinite momentum frame is singular at the boundary $V\sim m^2/X$. However, once the potential is regulated $X\rightarrow \sqrt{X^2+1/\gamma^2}$ via a finite gamma factor, the boundary term turns our to be numerically small, as it follows from the $squared$ small values tabulated above.

This procedure was extended  to the P-shell
of the pentaquarks. The shape of the function $\psi[\{2,1,1,1\}]$
(black solid line) is compared to the variational ones in Fig.\ref{fig_L1_var}. Note that  even a simple one-parameter  variational function (red dashed) takes the shape close to the exact one, except at the left side of the plot (where  $x_5\rightarrow 1$).
This observation is quite typical.
While the Ritz variational method can lead to
reasonable wave functions overall, their predictions of the tails are not reliable since they
play little role in the minimization of the integrals.

\begin{widetext}

The general formula for the eigenvalues on a unit $d$-simplex is
\be
    E(m_1,m_2,\dots m_{d-1})\propto\pi^2 |p|^2=
    \pi^2\sum_{i=1}^{d-1}\sum_{j=1}^{d-1}\left(\min(i,j)-\frac{ij}{d}\right)m_i m_j,
\ee
which for integer values, become
\ba E(m1,m2,m3,m4)\sim (2m_1^2+3m_2^2+3m_3^2+3m_3m_4+2m_4^2 \nonumber \\
+4m_2m_3+2m_2m_4+3m_1m_2+2m_1m_3+m_1m_4)\ea
with a common scaling factor from the coordinate transform from $x$ to $y$.

Therefore the ratio of the P-wave to the S-wave eigenvalues is predicted to be
\be {Pwave \over Swave}=(74/5)/10 =1.48\ee
This ratio is lower than the ratio of the values obtained
from (\ref{eqn_ansatz_faces}) prior to the variational corrections
\be {Pwave \over Swave}={312./182.}=1.71
\ee
The latters tend to lower this
ratio towards the exact eigenvalues, reaching about 1.6. 

The Laplacian eigenfunctions are smaller
near the corners (the left side of the plots versus $\delta$) than the variational ones.  If we  evaluate  the PDFs (the integral of the corresponding squared WF  over the three momentum fractions $\alpha,\beta,\gamma$), a logarithmic plot shows a   larger hard edge behavior  $\sim (1-x)^p, p\approx 12$. A imilar behavior was observed in~\cite{PENTAX}.

The averaging of the potential $V_{cup}$ leads to  smaller values for the Laplacian functions, in comparison to  the variational ones, to the extent
that it can even be neglected within a  10\% accuracy.
One may wander why $V_{cup}$ is more important for baryons (on a triangle or 2-simplex) compared to pentaquarks (on a pentachoron or  4-simplex). It maybe a general trend as $N$ increases.

\newpage

Finally and for definiteness, let us give the explicit form of the lowest Laplacian eigenstate $\psi(1,1,1,1)$. 
\begin{small}
\ba & Re& \psi[1,1,1,1]= \\
&\cos(&2.5651 \beta +13.3389 \gamma -1.5708 \delta ) 
  -\cos (-10.4036 \alpha +4.85939 \beta +7.37354 \gamma -0.868315
   \delta +1.25664)
   \nonumber \\
-&\cos(&-3.2149 \alpha -12.7221 \beta +3.74594 \gamma -0.868315 \delta +1.25664)
   +\cos (-8.41672 \alpha
   -9.71879 \beta +4.65284 \gamma -0.165833 \delta +2.51327)\nonumber \\
+&\cos(&-9.41018 \alpha -0.573574 \beta +9.90275 \gamma -0.165833
   \delta +2.51327)
   +\cos (9.41018 \alpha +7.99807 \beta +5.65549 \gamma +1.5708 \delta )\nonumber \\
-&\cos(&-5.20182 \alpha +0.438171
   \beta -12.432 \gamma +2.27328 \delta +1.25664)
   -\cos (-6.19528 \alpha +9.58339 \beta -7.18206 \gamma +2.27328 \delta
   +1.25664)\nonumber \\
-&\cos(&0.993459 \alpha -7.99807 \beta -10.8097 \gamma +2.27328 \delta +1.25664)
   -\cos (4.20836 \alpha +11.0013
   \beta +6.56239 \gamma +2.27328 \delta +1.25664)\nonumber \\
 -&\cos(&10.4036 \alpha +2.5651 \beta +8.1847 \gamma +2.27328 \delta
   +1.25664)
   -\cos (11.3971 \alpha -6.58012 \beta +2.93479 \gamma +2.27328 \delta +1.25664)\nonumber \\
 -&\cos(&9.41018 \alpha +5.43297
   \beta -7.68338 \gamma +3.14159 \delta )
   +\cos (-9.41018 \alpha -1.99153 \beta -8.99585 \gamma +3.67824 \delta +3.76991)\nonumber \\
+ &\cos(&-10.4036 \alpha +7.15369 \beta -3.74594 \gamma +3.67824 \delta +3.76991)+\cos (-3.2149 \alpha -10.4278 \beta -7.37354
   \gamma +3.67824 \delta +3.76991)\nonumber \\
+&\cos(&8.57164 \beta +9.9985 \gamma +3.67824 \delta +3.76991)+\cos (7.18874 \alpha
   -9.00981 \beta +6.3709 \gamma +3.67824 \delta +3.76991)\nonumber \\
+&\cos(&6.19528 \alpha +0.135402 \beta +11.6208 \gamma +3.67824
   \delta +3.76991)-\cos (5.20182 \alpha -0.708976 \beta -12.0264 \gamma +3.84407 \delta +1.25664)\nonumber \\
-&\cos(&9.41018 \alpha
   +9.14522 \beta +0.0957439 \gamma +3.84407 \delta +1.25664)-\cos (-8.41672 \alpha -7.42449 \beta -6.46664 \gamma +4.38072
   \delta +5.02655)\nonumber \\
-&\cos(&0.993459 \alpha +3.13867 \beta +12.5277 \gamma +4.38072 \delta +5.02655)+\cos (-0.993459 \alpha
   +11.4395 \beta -5.86958 \gamma +4.54656 \delta +2.51327)\nonumber \\
+&\cos(&6.19528 \alpha -6.14194 \beta -9.49718 \gamma +4.54656
   \delta +2.51327)+\cos (4.20836 \alpha +12.1485 \beta +1.00264 \gamma +4.54656 \delta +2.51327)\nonumber \\
+&\cos(&11.3971 \alpha
   -5.43297 \beta -2.62496 \gamma +4.54656 \delta +2.51327)+\cos (-9.41018 \alpha +5.43297 \beta +6.56239 \gamma +5.0832
   \delta +6.28319)\nonumber \\
+&\cos(&-2.22144 \alpha -12.1485 \beta +2.93479 \gamma +5.0832 \delta +6.28319)+\cos (-5.20182 \alpha
   +5.29757 \beta -10.2126 \gamma +5.24904 \delta +3.76991)\nonumber \\
+&\cos(&11.3971 \alpha -1.72072 \beta +5.15417 \gamma +5.24904
   \delta +3.76991)-\cos (-7.42326 \alpha -9.14522 \beta +3.84169 \gamma +5.78569 \delta +7.53982)\nonumber \\
-&\cos(&-8.41672 \alpha
   +9.0916 \gamma +5.78569 \delta +7.53982) -\cos (-9.41018 \alpha +2.86787 \beta -6.77648 \gamma +6.654 \delta +6.28319)\nonumber \\
   -
   &\cos(&7.18874 \alpha -4.15042 \beta +8.59028 \gamma +6.654 \delta +6.28319)+\cos (5.20182 \alpha +4.15042 \beta -9.80701
   \gamma +6.81983 \delta +3.76991)\nonumber \\
+&\cos(&10.4036 \alpha +4.85939 \beta -2.93479 \gamma +6.81983 \delta +3.76991)-\cos
   (-9.41018 \alpha +6.58012 \beta +1.00264 \gamma +7.35648 \delta +7.53982)\nonumber \\
-&\cos(&-2.22144 \alpha -11.0013 \beta -2.62496
   \gamma +7.35648 \delta +7.53982)-\cos (-4.20836 \alpha +7.28909 \beta +7.87487 \gamma +7.35648 \delta +7.53982)\nonumber \\
-&\cos&
   (2.98038 \alpha -10.2924 \beta +4.24727 \gamma +7.35648 \delta +7.53982)-\cos (7.15369 \beta -8.90011 \gamma +7.52231
   \delta +5.02655)\nonumber \\
-&\cos(&11.3971 \alpha -0.573574 \beta -0.405578 \gamma +7.52231 \delta +5.02655)+\cos (-7.42326 \alpha
   -7.99807 \beta -1.71806 \gamma +8.05896 \delta +8.79646)\nonumber \\
+&\cos(&-3.2149 \alpha +1.85612 \beta +10.4041 \gamma +8.05896
   \delta +8.79646)-\cos (10.8659 \beta -1.12099 \gamma +8.2248 \delta +6.28319)\nonumber \\
-&\cos(&7.18874 \alpha -6.71552 \beta
   -4.74859 \gamma +8.2248 \delta +6.28319) 
+\cos (-7.42326 \alpha -4.28582 \beta +6.06107 \gamma +8.76144 \delta +10.0531)\nonumber \\
+
   &\cos(&-4.20836 \alpha +8.43624 \beta +2.31512 \gamma +9.62976 \delta +8.79646)+\cos (2.98038 \alpha -9.14522 \beta -1.31248
   \gamma +9.62976 \delta +8.79646)\nonumber \\
   +&\cos(&-8.41672 \alpha +2.29429 \beta -2.02789 \gamma +10.3322 \delta +10.0531)+\cos
   (2.98038 \alpha -5.43297 \beta +6.46664 \gamma +10.3322 \delta +10.0531)\nonumber \\
-&\cos(&6.19528 \alpha +3.57684 \beta -5.05842
   \gamma +10.4981 \delta +7.53982)-\cos (-7.42326 \alpha -3.13867 \beta +0.501322 \gamma +11.0347 \delta +11.3097)\nonumber \\
-&\cos&
   (-2.22144 \alpha -2.4297 \beta +7.37354 \gamma +11.0347 \delta +11.3097)+\cos (0.993459 \alpha +6.58012 \beta -4.15152
   \gamma +11.2006 \delta +8.79646)\nonumber \\
+&\cos(&7.18874 \alpha -1.85612 \beta -2.52921 \gamma +11.2006 \delta +8.79646)-\cos
   (-3.2149 \alpha +4.15042 \beta -0.715412 \gamma +12.6055 \delta +11.3097) \nonumber \\
   -&\cos(& 2.98038 \alpha -4.28582 \beta +0.9069
   \gamma +12.6055 \delta +11.3097) +\cos(-2.22144 \alpha -1.28255 \beta +1.8138 \gamma +13.308 \delta +12.5664)) \nonumber
\ea
\end{small} 
\end{widetext}

\section{Large $N$ and the ``spherical approximation"} \label{sec_large_N}
If the number of corners $N$ is large, the corresponding simplexes are bracketed by spheres. The outer or ``large" sphere with the radius at the corners or
$R_{large}=| a_i |$, and theinner or  "small" sphere with  the radius normal to the edges. This spherical approximation maybe useful for
higher Fock components,  but for now $N=5$ is enough
to provide the first glimpse for the antiquarks.

For $N=5$ the number of edges, 10, can be considered large. The numerical values of the radii are
\ba R_{\rm large}&=&{2 \over \sqrt{5}}\approx 0.894,\nonumber \\
R_{\rm small}&=&\sqrt{3 \over 10}\approx 0.548 
\ea
which are not that close, but they should become  closer at larger $N$

The radial Schrodinger equation for 4d spheres, is 
of the form
\be -u''(r) +\big({3 \over 4 r^2}+V(r)\big)u(r)  \ee
where $u=\psi*r^{3/2}$. Without potential (pure Laplacian) the ground state
eigenvalues are 18.353  for our large sphere, and 48.94 for the
small sphere. With the potential $3/(R-r)$ (imitating our $V_{cup}$), 
those values increase substantially, to 27.10 and 63.51
respectively. The values for our variational ansatz $\Psi_{0.8}$ are larger still, about
70 and 150, respectively.

\begin{figure}
    \centering
    \includegraphics[width=0.85\linewidth]{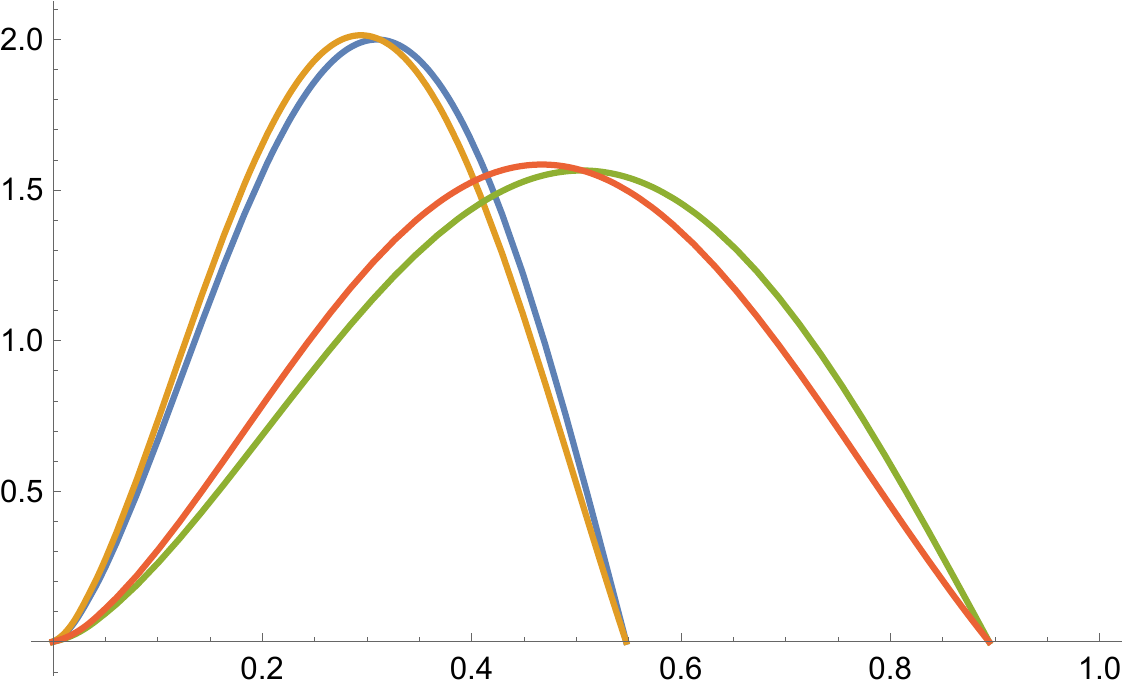}
    \caption{The ground state wave functions $u(r)$, for the large and small spheres, with (red and orange) and without the cup potential (blue and green).}
    \label{fig_wf_spheres}
\end{figure}

Yet, as 
Fig.\ref{fig_wf_spheres} shows, the wave functions are
hardly modified by the potential. Perhaps this is also the case for the actual 4-simplex.

\bibliography{penta2}
\end{document}